\documentclass[11pt,letter]{article}
\pdfoutput=1 %% activate this line when submit %%

\usepackage{jheppub}
\usepackage{amssymb,amsmath,amsthm,graphicx}
\usepackage{graphics, color}
\usepackage{subfigure}
\usepackage{latexsym}
\usepackage{bm}
\usepackage[hang,flushmargin]{footmisc}
\usepackage{epsfig}
\usepackage{textcomp}
\usepackage{cancel}
\hypersetup{
 pdfauthor={Gary T. Horowitz, Donald Marolf, Jorge  E. Santos},
 citecolor=black,linkcolor=black,urlcolor=black}
\usepackage[utf8]{inputenc}
\allowdisplaybreaks[1]

\newcommand{\RED}{\color{red}}
\newcommand{\be}{\begin{equation}}
\newcommand{\ee}{\end{equation}}

\begin{document}
%---Title and Abstract -----------------------------------------------------------
\title{Constraints are not enough}

\author[1]{Gary~T.~Horowitz,}
\author[1]{Donald~Marolf}
\author[2]{and Jorge~E.~Santos}

\affiliation[1]{Department of Physics, University of California, Santa Barbara, CA 93106, U.S.A.}
\affiliation[2]{DAMTP, Centre for Mathematical Sciences, University of Cambridge, Wilberforce Road, \\ Cambridge CB3 0WA, UK}
\emailAdd{horowitz@ucsb.edu}
\emailAdd{marolf@ucsb.edu}
\emailAdd{jss55@cam.ac.uk}

\newcommand{\blue}{\color{blue}}
\newcommand{\red}{\color{red}}

\abstract{The Euclidean Einstein-Hilbert action is well-known to be unbounded below and thus to raise many questions regarding the definition of the gravitational path integral.  A variety of works since the late 1980's have suggested that this problem disappears when one fixes a foliation of the spacetime and imposes the corresponding gravitational constraints. However, we show here that this approach fails with various classes of boundary conditions imposed on the foliation: compact slices without boundary, asymptotically flat, or asymptotically locally anti-de Sitter slices.  We also discuss the idea of fixing the scalar curvature and Wick-rotating the conformal factor, and show that it also fails to produce an action bounded from below.} 
%in light of recent mathematical results.}

\maketitle
\section{Introduction}

Euclidean path integrals are a primary tool in the study of quantum gravity.  They play critical roles in computing black hole entropy \cite{Gibbons:1976ue}, tunneling from metastable vaccua \cite{Coleman:1980aw,Witten:1981gj}, the Page curve from replica wormholes \cite{Penington:2019kki,Almheiri:2019qdq}, Schwarzian-mode contributions to near-extremal black hole entropy \cite{Iliesiu:2020qvm,Heydeman:2020hhw}, and recent discussions of phases asociated with de Sitter partition functions \cite{Maldacena:2024spf,Ivo:2025yek,Shi:2025amq}. 

Despite these successes, such integrals suffer from the fact that the Euclidean gravitational action can be arbitrarily negative, even with fixed boundary conditions \cite{Gibbons:1978ac}.  The integral over all real Euclidean geometries thus appears to diverge and cannot be taken to define the theory.  The problem persists even at the semiclassical level as, without knowing the fundamental definition of the proper contour of integration, it is impossible to fully determine which saddles contribute in the semiclassical limit.   We also emphasize that, as we review in Section \ref{sec:compact} below, the fact that the action is unbounded below manifests itself not only in the UV but also in the IR so that it seems unlikely to be resolved by short-distance corrections.

There are suggestions in the literature that negative Euclidean actions are a result of degrees of freedom that violate the gravitational constraints.    Such suggestions date back at least to the 1987 work of Hartle and Schleich \cite{Hartle:2020glw,Schleich:1987fm} which showed that, given the standard foliation of Minkowski space by constant time slices defined in some inertial frame, the Euclidean action for linearized gravity in flat Euclidean space becomes positive definite when the linearized constraints hold on each slice.  Here the term constraints is used to describe equations of motion that arise in the 2nd-order formalism that restrict the allowed Cauchy data; i.e.,  which contain at most first derivatives normal to each slice\footnote{In contrast, in the canonical formalism the off-shell action is a function of independent choices of lapse, shift, momenta, and induced metrics on each slice.  In particular, at this stage no relation is imposed between the momenta and extrinsic curvatures.  As a result, enforcing the associated constraints and taking the Euclidean versions of the above quantities to be real is then insufficient to make the Euclidean action bounded below, even in the linearized theory.  This is analogous to the fact that the off-shell Euclidean action $S_E  = \int d\tau (p_E\dot{q}+H_E)$  is unbounded below even for a simple Harmonic oscillator (for which $H_E = -\frac{p_E^2}{2m} + \frac{1}{2}kq^2$).  In constrast, imposing the equation of motion $p_E = \dot{q}/2m$ gives the positive-definite result $S_E = \frac{1}{2}\int dt_E (m\dot{q}^2 + kq^2)$.}.  As described in appendix \ref{app:poslin}, the Hartle-Schleich argument generalizes readily to linearized gravity with a negative cosmological constant using the the global or Poincar\'e foliation of anti de-Sitter space.   We also take the opportunity in appendix \ref{app:poslin} to transcribe the argument into a form that is more transparent for our current purposes.   A similar result was also established at the non-perturbative level by Hajicek \cite{Hajicek:1984nb} for a toy model of Euclidean gravity.  Comments in \cite{Kuchar:1970mu,Dasgupta:2001ue,Mazur:1989by} have also been interpreted as supporting similar conclusions in Einstein-Hilbert gravity.  As a result, some recent papers \cite{Banihashemi:2022htw,Banihashemi:2024weu} have taken as fact the claim that the Euclidean Einstein-Hilbert action becomes bounded below when one imposes the constraints defined by a smooth foliation.

 We show below that this is simply not the case for a variety of common boundary conditions including cosmology
(i.e., taking the slice of the foliation to be compact manifolds without boundary), asymptotically anti-de Sitter (AdS) slices, or asymptotically flat slices.  We emphasize that our counterexamples are all smooth, so that there is no ambiguity in interpreting the constraints\footnote{In contrast, this project was inspired by forthcoming work \cite{toappear} that considered off-shell configurations that are continuous, but not differentiable, on certain codimension-1 surfaces where it is more subtle to evaluate whether or not the constraints in fact hold.}.  One hint that the Euclidean constraints do not ensure positive action is that terms involving the square of the extrinsic curvature $K_{ab}$ (or time derivatives of matter fields) appear with the opposite sign compared to the Lorentzian constraints. For example, the Euclidean  constraints are 
\begin{align}
\label{eq:EHcons}
{\cal R} +K_{ab}K^{ab} - K^2 & = 0\\
\label{eq:EPcons}
D^a (K_{ab} - K h_{ab}) & = 0
\end{align}
where ${\cal R}$ is the scalar curvature of the induced metric on the slice $h_{ab}$. When these constraints are satisfied on a family of slices foliating the space, the gravitational action 
\be
S_E=-\frac{1}{16\pi}\int_{\mathcal{M}} \,R\,\sqrt{g}\,{\rm d}^{D}x -\frac{1}{8 \pi }\oint_{\partial \mathcal{M}} \tilde{K}\,\sqrt{\tilde h}\,{\rm d}^{D-1}x + S_{ct}
\ee
(where  $\tilde{K}$ is the trace of the extrinsic curvature of the boundary with metric $\tilde  h_{ab}$, $S_{ct}$ denotes possible counterterms needed to render the action finite, and we have set $G=1$)
reduces to 
\be
\label{eq:EHaction}
S_E = \frac{1}{8\pi}\int (K_{ab}K^{ab} -K^2) \sqrt{g}\ {\rm d}\tau\,{\rm d}^{D-1} x + \int M {\rm d}\tau.
\ee
 where $M$ denotes the total mass of the spacetime. The positive energy theorem in general relativity shows that $M \ge 0$ if the Lorentzian constraints are satisfied. But since \eqref{eq:EHcons} has changed the sign of the extrinsic curvature terms, $M$ can now  be arbitrarily negative even when the Euclidean constraints are satisfied.

%Due to the above signs, the total mass, $M$ can now  be arbitrarily negative for real Euclidean-signature asymptotically-flat or asymptotically-AdS spacetimes, even when \eqref{eq:EHcons} and \eqref{eq:EPcons} are satisfied. 

Echos of the above discussion also infect what might at first seem to be more prosaic attempts to study quantum gravity perturbatively about Euclidean saddles.  At issue here is the question of whether a given Euclidean saddle contributes to the $G\rightarrow 0$ 
 asymptotic expansion of a given partition function, a question often described as investigating the `stability' of the saddle or checking for the existence of `negative modes'.  From a fundamental perspective, finding a solid answer to this question would first require having a sufficient definition of the theory.  In the present context, this would naturally be taken to be the specification of a contour through the space of complex metrics that can be used to define the `Euclidean' path integral.  However, in the absence of such a specification it has become a common practice to follow  \cite{Garriga:1997wz,Gratton:1999ya,Kol:2006ga}
(see e.g. \cite{Monteiro:2008wr,Hertog:2018kbz,Marolf:2021kjc,Loges:2022nuw,Loges:2023ypl,Aguilar-Gutierrez:2023ril,Hertog:2024nys} for other works using this procedure) and simply linearize the theory about the saddle in some natural foliation, perform whatever Wick rotations are required to semiclassically evaluate the integrals that impose the gravitational constraints (and also the momentum integrals in a canonical approach), and then to carefully study the resulting quadratic action $\tilde S^{(2)}$ as a functional of real Euclidean fluctuations about the saddle.  If and only if this $\tilde S^{(2)}$ is positive definite is the saddle then deemed `stable' in this approach and taken to contribute to the asymptotic expansion of the partition function.   

While we can find no explicit comment in the literature tying the procedure of \cite{Garriga:1997wz,Gratton:1999ya,Kol:2006ga}  to  discussions of whether Euclidean actions in the full nonlinear theory are bounded below, we note that the above $\tilde S^{(2)}$  is precisely the covariant gravitational action on the constraint surface defined by the foliation.  A result that the full non-linear action was indeed bounded below on the constraint surface would thus help to justify the above approach to semiclassical stability.  Furthermore, in the absence of such a result the above perturbative procedure seems rather ad hoc\footnote{This does not necessarily imply that the procedure gives incorrect answers.  Indeed, we know of no physical problem or sharp contradiction that would arise from claiming that positivity of $\tilde S^{(2)}$ about a saddle is the correct prescription for determining its relevance.  However, since the full non-linear action is unbounded below on the constraint surface, we have no idea how such a perturbative prescription might be derived.}.

 In addition to discussing the action on the constraint surface, we will also briefly consider an alternative approach to defining the path integral originally suggested by Gibbons, Hawking and Perry \cite{Gibbons:1978ac} for asymptotically Euclidean metrics.  They suggested that one write the full metric $g$ in the form $\Omega^2 \tilde g$ where $\tilde g$ has zero scalar curvature. This is possible for all metrics for which the lowest eigenvalue of the conformally invariant Laplacian (with the field kept constant on the boundary) is positive.  They made a ``positive action" conjecture which stated that the action for all asymptotically Euclidean metrics with $R = 0$ was nonnegative. If true, they argued that one could obtain a positive action for all conformally related metrics by simply Wick-rotating the conformal factor.  This positive action conjecture is a higher dimensional analog of the positive energy conjecture and was proven by Schoen and Yau \cite{SY}. However, this prescription only applied to the subset of metrics that could be rescaled to have $R = 0$, and no analogous proposal was given for the rest, so it was unsatisfactory. 

We consider the analogous prescription for negative cosmological constant $\Lambda$. Namely, we write the full metric $g$ in the form $\Omega^2 \tilde g$ where $\tilde g$ has constant negative scalar curvature. This has the immediate advantage that all asymptotically hyperbolic Riemannian manifolds can be rescaled to have constant negative $R$ \cite{Andersson:1992yk,allen2022}. Unfortunately, we find that there is no analog of the positive action theorem in this case: including a matter field allowed by supergravity, we find configurations with constant negative $R$ and arbitrarily negative action.

Below, we provide various examples of families of real Euclidean configurations for which the Euclidean gravitational action is unbounded below. (We restrict throughout to connected manifolds.)  None of our examples will satisfy the full set of Einstein's equations. In sections 2 - 4, the examples satisfy the Euclidean constraints, and in sections 5 and 6 they have constant Ricci scalar {but do not otherwise satisfy the equations of motion}.  Section \ref{sec:compact} provides examples satisfying cosmological boundary conditions, by which we mean that each slice $\Sigma$ in the given foliation is taken to be a closed manifold (without boundary).    We exhibit such families for negative, positive, or zero cosmological constant.   Section \ref{sec:asymflat} then presents examples that are asymptotically Euclidean or asymptotically flat with thermal boundary conditions. In
Section \ref{sec:AdS} we give examples that satisfy asymptotically locally AdS boundary conditions and, in particular, which are asymptotic to thermal AdS.   We then turn in Section \ref{sec:constR} to fixing the Ricci scalar instead of imposing the constraints and show that this does not provide a lower bound to the action. In Section \ref{sec:complex} we show that, if one allows the type of complex metrics that arise from analytically continuing a real Lorentzian rotating solution, one can also easily obtain configurations with constant $R < 0$ and unbounded action.   We conclude with some final discussion in Section \ref{sec:disc}.

%\newpage
\section{The case of compact slices}
\label{sec:compact}

The fact that the Euclidean action is unbounded below can famously be traced to a certain sign in the kinetic term.  This sign can also be seen in the final $K^2$ term of the kinetic term in the action \eqref{eq:EHaction}, which comes with a sign opposite to the rest of the kinetic term.  Since $K$ is related to the time-derivative of the spatial volume element, the issue is often called the conformal factor problem.  

The above negative contribution to the kinetic term means that $S_E$ becomes large and negative when the induced metric on any slices $\Sigma$ varies only by a Weyl rescaling, and where the corresponding conformal factor $\Omega$ varies rapidly.  For this reason, one might think that this issue is entirely a UV problem that should be resolved by an appropriate UV completion.  

However, the problem also manifests itself as a purely infrared issue.  To see this, it is useful to begin by considering the case where the full $D$-dimensional spacetime is a $D$-sphere of radius $r_0$.  The Einstein-Hilbert term $-\frac{1}{16\pi}\int \sqrt{g}\,R\,{\rm d}^Dx$ in the action is then negative.  A negative cosmological constant will only make the action more negative and, in the case of a positive cosmological constant, we can still make the full action negative by taking $r_0$ to be sufficiently small. 

\begin{figure}[tb]
    \centering
    \includegraphics[width=0.6\textwidth]{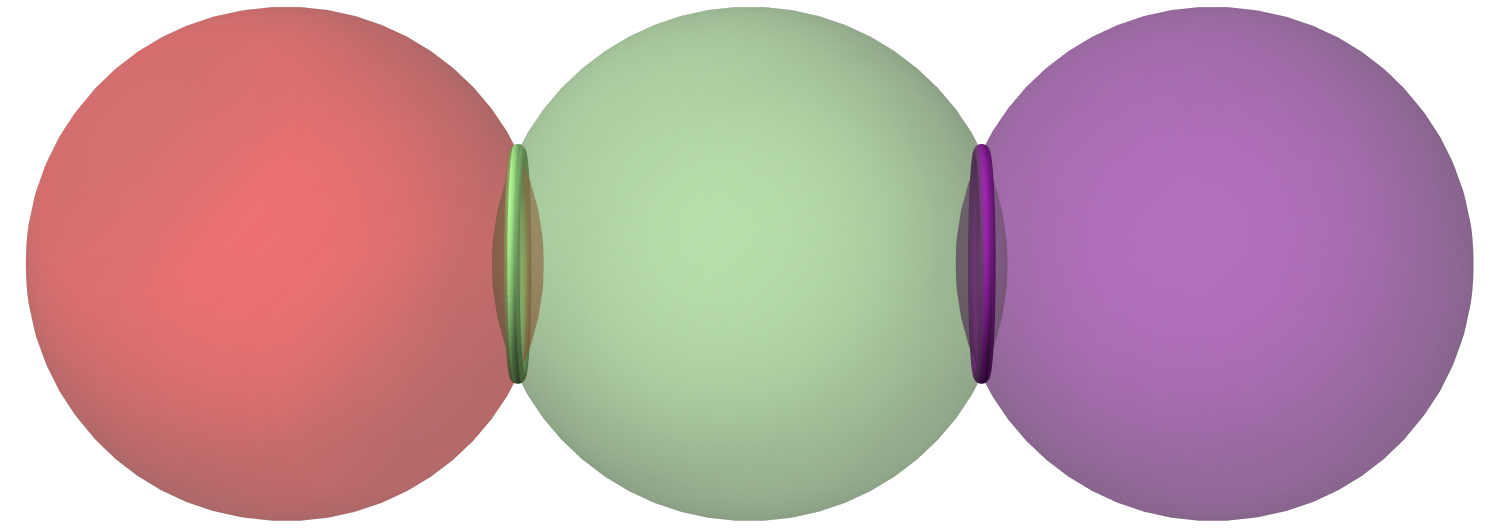}
    \caption{Gluing three four-spheres to make the  Euclidean action more negative.  Gluing additional spheres can make the action arbitrarily negative.}
    \label{fig:illu}
\end{figure}

The disconnected union of many such spheres then has arbitrarily negative action.  Furthermore, by cutting holes of some small radius $\epsilon$ in each sphere near the poles and gluing the remainders of the spheres together in a chain, we can construct a connected manifold with arbitrarily negative action. See Fig.~\ref{fig:illu} for an illustration of this construction.
 For spacetime dimensions $D\ge2$, any positive contribution from the seams where two spheres are joined scales to zero with a positive power of $\epsilon$ and gives only a small correction to the action of the disjoint union.   
Indeed, the action remains negative and linear in the number of spheres even when the seams are smoothed out and when $\epsilon/r_0$ is a small number of order, say, $.1$.  We thus see that large negative actions can arise due to IR effects; there is no need for curvatures to become parametrically large.

Below, we show that similar families of spacetimes exist even when the constraints \eqref{eq:EHcons} and \eqref{eq:EPcons} are imposed on each slice.  We begin with the case $\Lambda >0$ in section \ref{sec:posLambda} and then address $\Lambda \le 0$ in section \ref{sec:nonposLambda}.

\subsection{Positive cosmological constant}
\label{sec:posLambda}

In this section we show that one can
 join together a series of spheres in a way that satisfies the constraints on each slice $\Sigma$.  To do so, it is convenient to again take each $\Sigma$ to be topologically a $(D-1)$-sphere, but to allow the metric on each slice to be somewhat squashed.  In particular, let us consider the case $D=4$ and take the spacetime to have the form of a so-called locally rotationally symmetric Bianchi IX cosmological model (see e.g. \cite{Misner:1973prb} with $\beta_-=0$).  We write the spacetime metric in the form
\begin{equation}
{\rm d}s^2={\rm d}\tau^2+\frac{a(\tau)^2}{4} \left[e^{-2 B(\tau)}\sigma_3^2+e^{B(\tau)}\left(\sigma_1^2+\sigma_2^2\right)\right]\,,
\end{equation}
with
\begin{align}\label{eq:sigmas}
&\sigma_1=-\sin 2\psi\,{\rm d}\theta+\cos 2\psi\,\sin\theta\,{\rm d}\phi \nonumber
\\
&\sigma_2=\cos 2\psi\,{\rm d}\theta+\sin 2\psi\,\sin\theta\,{\rm d}\phi
\\
& \sigma_3=2{\rm d}\psi+\cos \theta\,{\rm d}\phi\nonumber
\end{align}
with $\phi\sim\phi+2\pi$, $\psi\sim\psi+2\pi$ and $\theta\in[0,\pi]$. When $B=0$, the metric at each $\tau$ is a round $S^3$.

With this ansatz, the momentum constraint is satisfied identically and the Hamiltonian constraint reduces to the single equation
\begin{equation}
\label{eq:HLRSBIX}
n^a n^bE_{ab}=0\Rightarrow \frac{3 \dot{a}(\tau )^2}{a(\tau )^2}+\frac{3}{L^2}-\frac{3}{4} \dot{B}(\tau )^2+\frac{e^{-4 B(\tau )}-4 e^{-B(\tau )}}{a(\tau )^2}=0,
\end{equation}
where $\dot{\,}$ represents a derivative with respect to $\tau$ and we have used
\begin{equation}
n_a n^a=1\,\quad\text{and}\quad E_{ab}=R_{ab}-\frac{R}{2}g_{ab}+\Lambda\,g_{ab},\quad\text{with}\quad \Lambda=\frac{3}{L^2}\,.
\end{equation}

Let us now restrict attention to spacetimes that satisfy
\begin{equation}
\label{eq:Bconst}
B(\tau)=\varepsilon\,\log\left[\frac{a(\tau)}{L}\right],
\end{equation}
and let us introduce the dimensionless variable
\begin{equation}
A\left(\frac{\tau}{L}\right) = a(\tau)/L\,\,.\nonumber
\end{equation}
The Hamiltonian constraint then reduces to
\begin{equation}
\left(1-\frac{\varepsilon ^2}{4}\right) A'(x)^2= - A(x)^2 - \frac{1}{3 A(x)^{4 \varepsilon }}+ \frac{4}{3 A(x)^{\varepsilon }}=0\,,\label{eq:C1}
\end{equation}
where $x\equiv \tau/L$ and ${}^\prime$ represents derivative with respect to $x$.
We wish to study solutions to this constraint.

%that are in fact solutions of the constrained problem defined by \eqref{eq:action}; i.e., which satisfy the equations of motion obtained by inserting \eqref{eq:action} into the gravitational action.  These spacetimes are off-shell with respect to the original problem, but they will satisfy the constraint \eqref{eq:C1} on each slice.

For $\varepsilon=0$, one finds
\begin{equation}
A=\sin\left(x+x_0\right),
\end{equation}
representing the usual round Euclidean 4-sphere. However, the behavior is more interesting for $0<\varepsilon < 2$.  Since the left-hand side of \eqref{eq:C1} is positive definite,  the constraint then requires the final term on the right-hand side to be larger than the others.  This condition then 
imposes a lower bound on $A(x)$ since the 2nd term on the right-hand side would clearly dominate in the limit $A(x)\rightarrow 0$.  In addition, it also imposes an upper bound since the first term on the right clearly dominates in the limit $A(x) \rightarrow +\infty$.  

Indeed, it is easy to check that the right-hand side of \eqref{eq:C1} vanishes  at $A(x)=1$ for any value of $\varepsilon$.  It is also straightforward to check that this defines an
upper turning point, which we call $A^+$.   For any $0<\varepsilon<2$, there is \emph{another} turning point, $A^-$, which for small enough $\varepsilon$ is well approximated by
\begin{equation}
A^-\approx 2^{-\frac{2}{3 \varepsilon }}.
\end{equation}
For instance, for $\varepsilon=10^{-1}$ this gives
\begin{equation}
A^-\approx0.009844635(8)\,.
\end{equation}

\begin{figure}[tb]
    \centering
    \includegraphics[width=0.9\textwidth]{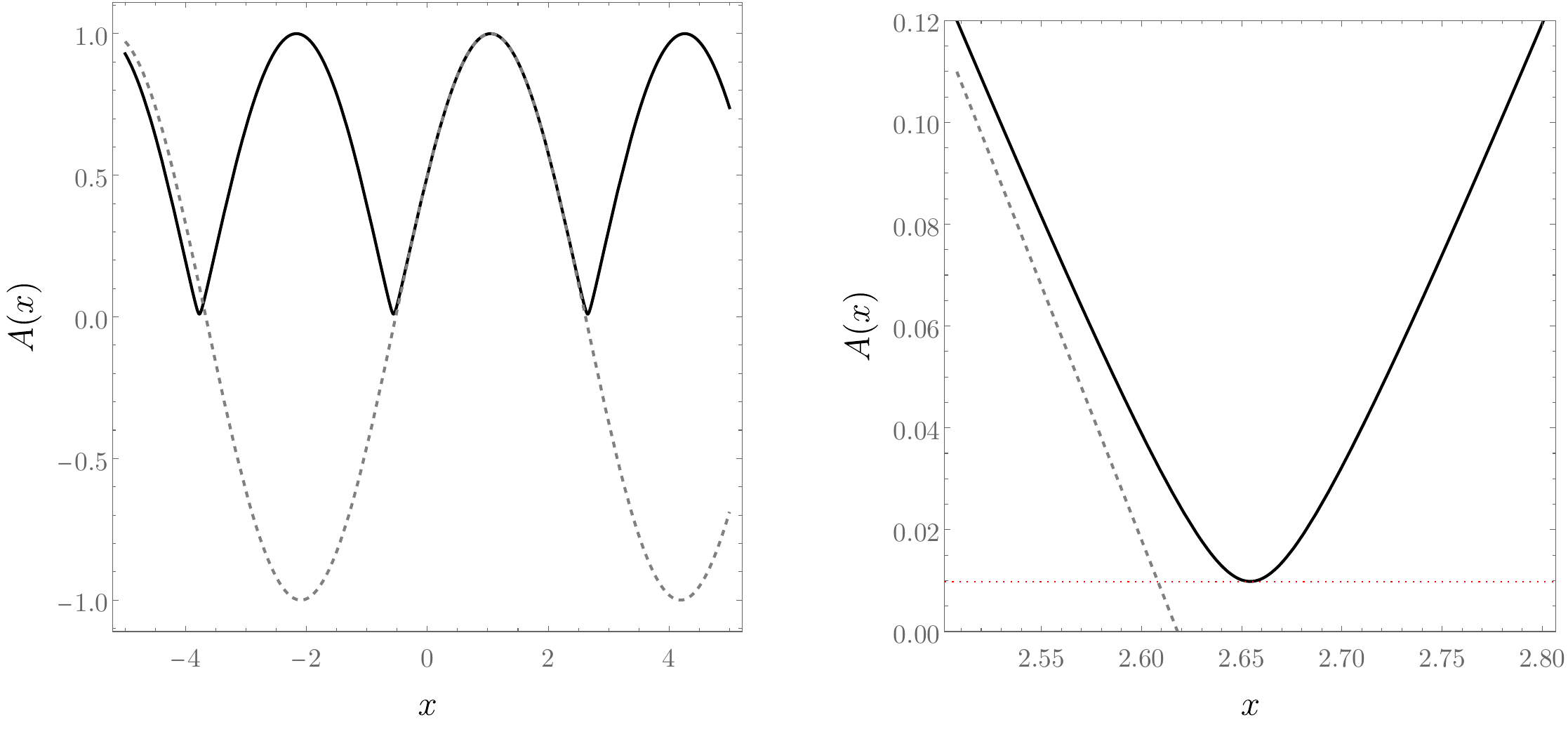}
    \caption{The black solid line shows $A(x)$ as a function of $x$ for $\varepsilon=10^{-1}$, while the dashed gray line shows the Euclidean four-sphere. The right panel zooms in on a region near where the function $A(x)$ bounces off the lower turning point $A^-$, with the horizontal dotted red line showing $A=A^-$.}
    \label{fig:one}
\end{figure}

Thus the motion is confined to a finite interval of positive $A$-values and any solution simply executes periodic oscillations in $x$.  
An example is shown in  in Fig.~\ref{fig:one}, where we plot $A(x)$ (solid black line) as a function of $x$ for $\varepsilon=10^{-1}$. To aid vizualisation we also plot $A(x)$ for a Euclidean four-sphere (dashed gray line). The right panel zooms in on a region near where the function $A(x)$ bounces off the lower turning point $A^-$, whose value is indicated by the horizontal dotted red line.

Furthermore, for $\varepsilon$ small the spacetime is very close to being a round 4-sphere except where $A$ is exponentially small; i.e., when $A$ is close to its lower turning point.  The resulting spacetime is thus again essentially a sequence of 4-spheres of radius $L$ joined together by thin necks.  As in our earlier discussion of constraint-violating configurations, since we are in $D>2$ spacetime dimensions the action of the thin necks is small for small $\varepsilon$.  Periodically identifying the solution after $n$ oscillations thus yields $S_E= nS_E^0(\varepsilon)$ where $n$ is the number of oscillations and $S_E^0(\varepsilon)$ approaches the (negative) action of the round 4-sphere as $\varepsilon\rightarrow 0.$  

\subsection{Negative or zero cosmological constant}
\label{sec:nonposLambda}

We can also use the locally-rotationally symmetric Bianchi IX model (again in $D=4$ spacetime dimensions) to provide examples of arbitrarily negative actions with compact slices $\Sigma$ for 
negative or zero cosmological constant.  In such cases, the Hamiltonian constraint becomes
\begin{equation}
\label{eq:HLRSBIXneg}
n^a n^bE_{ab}=0\Rightarrow \frac{3 \dot{a}(\tau )^2}{a(\tau )^2}-\frac{3}{L^2}-\frac{3}{4} \dot{B}(\tau )^2+\frac{e^{-4 B(\tau )}-4 e^{-B(\tau )}}{a(\tau )^2}=0,
\end{equation}
which differs from \eqref{eq:HLRSBIX} only by the sign in the second term, and where now $\Lambda = -\frac{3}{L^2}$ (with perhaps $L=\infty$).
Our strategy will again be to impose a relation between $A$ and $B$, though the desired relation will be a bit more complicated. 

To find the desired relation, let us write
\begin{equation}
\alpha\left(\frac{\tau}{L}\right) = \log\left[\frac{a(\tau)}{L}\right]\,,\quad \beta\left(\frac{\tau}{L}\right) = \frac{B\left(\tau\right)}{2}\,.
\end{equation}
and introduce the variable $x\equiv \tau/L$. The Hamiltonian constraint becomes
\begin{equation}
\label{eq:HCagain}
3 e^{2 \alpha} \left(-\alpha'^2+\beta'^2\right)=-3 e^{2 \alpha}\kappa-4 e^{-2 \beta}+e^{-8 \beta}\,,
\end{equation}
where $'$ represents a derivative with respect to $x$. We have introduced a bookkeeping parameter $\kappa$, where $\kappa = 0$ corresponds to a vanishing cosmological constant, and $\kappa = 1$ corresponds to a negative cosmological constant. For $\kappa = 0$, the parameter $L$ in the relation between $\alpha$ and $a$ is taken to be any arbitrary scale. 

Let us give the name $V(\alpha, \beta) = -3e^{2\alpha}\kappa + W(\beta)$ to the right-hand side of \eqref{eq:HCagain}.  In order to determine the right relation to impose, it is useful to study the curve defined by $V=0$.  Here we consider only the expression for $V$ given by the right-hand side of \eqref{eq:HCagain} without imposing the constraint itself.  In doing so, we see that $V$ can vanish only when $W$ is positive.  This requires $\beta < \beta_0 \equiv -\frac{1}{3}\log 2$. For $\beta$ near $\beta_0$, imposing $V=0$ gives
\begin{equation}
\label{eq:V0}
3e^{2\alpha}\kappa = \left.\frac{\rm{d}W}{{\rm d}\beta}\right|_{\beta=\beta_0} (\beta-\beta_0) + \dots,
\end{equation}
where we note that $\left.\frac{\rm{d}W}{{\rm d}\beta}\right|_{\beta=\beta_0}<0$. 
In particular, $\frac{{\rm d}\beta}{{\rm d}\alpha}$  along this curve is very small for $\beta$ near $\beta_0$. As a result, if -- again without imposing the constraint \eqref{eq:HCagain} -- one now computes the left-hand side of \eqref{eq:HCagain} along the curve $V=0$, a negative result is obtained in this region.  We may thus call this curve timelike with respect to the metric defined by  the left-hand-side of \eqref{eq:HCagain}.  

 Since there is only one constraint and the metric involves two functions, we are free to impose a relation between $\alpha$ and  $\beta$.  Suppose that we pick a relation that restricts $\alpha, \beta$ to some other timelike curve $\gamma$ in the $\alpha, \beta$ plane.  Enforcing the constraint then determines the motion along $\gamma$ uniquely. Since $\gamma$ is timelike, the left-hand side of \eqref{eq:HCagain} will be negative, so the motion along $\gamma$ will be confined to a region where $V$ is either negative or zero.  In particular, any point where $V=0$ on $\gamma$ 
will be a turning point so long as the  tangents to $\gamma$ and $V=0$ are not degenerate at that point.  We need therefore only engineer $\gamma$ to cross the locus $V=0$ in the above non-degenerate sense at least twice near the region  described by \eqref{eq:V0}, with $V<0$ on the segment between the two crossings\footnote{Had we chosen $\gamma$ to coincide with the curve $V=0$, our constraint would require both $\alpha$ and $\beta$ to be time-independent and the action would vanish.}.  This will then guarantee that the model constrained to $\gamma$ has solutions that oscillate periodically. 

The action is thus clearly proportional to the number oscillations $n$.  We now show that the coefficient is negative. The action, written in first order form, reads
\begin{eqnarray}\label{eq:compactact}
S_{E}&=&\frac{L^2\,\pi\,n}{4} \int_{0}^{X}e^{\alpha} \left[e^{-8 \beta}-4 e^{-2 \beta}-3 e^{2 \alpha} \left(\kappa +\alpha'^2-\beta'^2\right)\right]{\rm d}x\, \cr
&=&\frac{L^2\,\pi\,n}{2} \int_{0}^{X}e^{\alpha} \left[-3 e^{2 \alpha} \left(\alpha'^2-\beta'^2\right)\right]{\rm d}x\,
,
\end{eqnarray}
where $\alpha(x+X)=\alpha(x)$, and similarly for $\beta$.  In passing from the first to the second line we have used 
the constraint \eqref{eq:HCagain}  to write the potential terms in \eqref{eq:compactact} in terms of the kinetic terms.   Since our spacetime  has $|{\beta}'| \ll |{\alpha}'|,$ the kinetic term is dominated by the conformal mode and the final expression for the action is negative.

\begin{figure}[tb]
    \centering
    \includegraphics[width=0.6\textwidth]{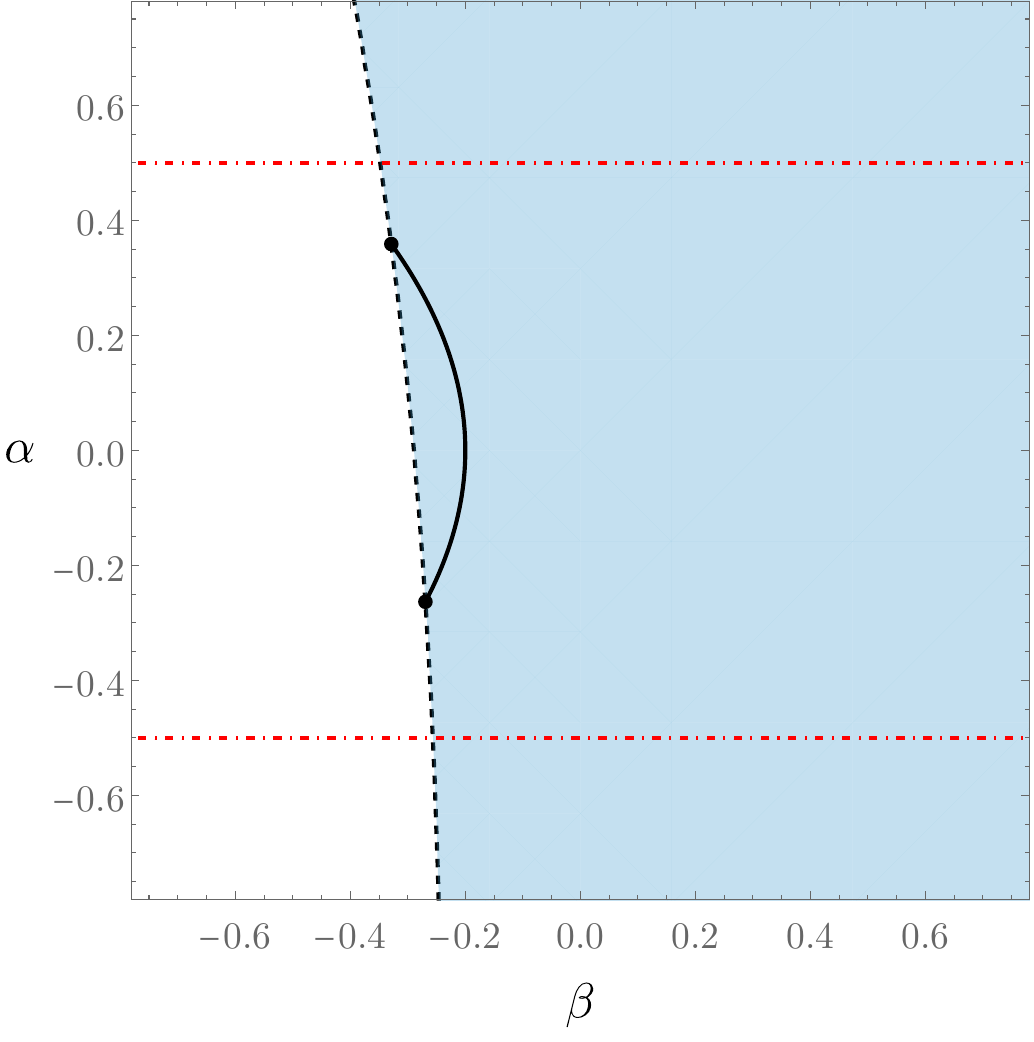}
    \caption{The $(\beta,\alpha)$ plane with the region $V < 0$ shown in blue. The black dashed line marks the boundary of this region. Horizontal dotted red lines indicate $|\alpha| = 1/2$ and we take $\kappa=1$. The trajectory $\gamma$ for $\eta = -0.2$, as given in Eq.~(\ref{eq:traalphabeta}), lies entirely within the allowed region.}
    \label{fig:moduli}
\end{figure}

To conclude this section, we present a concrete example and simultaneously address a more subtle point concerning the choice of $\gamma$. For $\gamma$, we take
\begin{equation}
\beta = \eta - \alpha^2\,,
\label{eq:traalphabeta}
\end{equation}
where $\eta$ is a real constant and we choose $\kappa=1$, i.e. negative cosmological constant. Substituting this expression into the constraint equation Eq.~(\ref{eq:HCagain}), we observe that the left-hand side becomes proportional to
\begin{equation}
3e^{2\alpha}{\alpha'}^2(4\alpha^2-1)=V(\alpha,\eta - \alpha^2)\,,
\end{equation}
which indicates that we require $|\alpha| < 1/2$ for our argument about the negativity of the right-hand side of Eq.~(\ref{eq:HCagain}) to hold. In Fig.~\ref{fig:moduli}, we display the $(\beta,\alpha)$ plane. The blue shaded region corresponds to $V < 0$, with its boundary marked by a black dashed line. Although we are now a finite distance away from $\beta_0$, so that we cannot blindly use \eqref{eq:V0}, it is clear from the figure that the numerically-determined black dashed line remains timelike. 

The horizontal dotted red lines indicate the boundaries at $|\alpha| = 1/2$. Our chosen trajectory $\gamma$, defined in Eq.~(\ref{eq:traalphabeta}), must be carefully chosen so that it both lies within the blue region and satisfies $|\alpha| < 1/2$. The first requirement imposes $\eta \gtrsim -0.289529(3)$, while the second requires $\eta \lesssim -0.0979(5)$. In Fig.~\ref{fig:moduli}, we take $\eta = -0.2$. This choice satisfies both conditions and yields a minimum at $\alpha \approx -0.26336(3)$ and a maximum at $\alpha \approx 0.35872(5)$.

\begin{figure}[tb]
    \centering
    \includegraphics[width=0.6\textwidth]{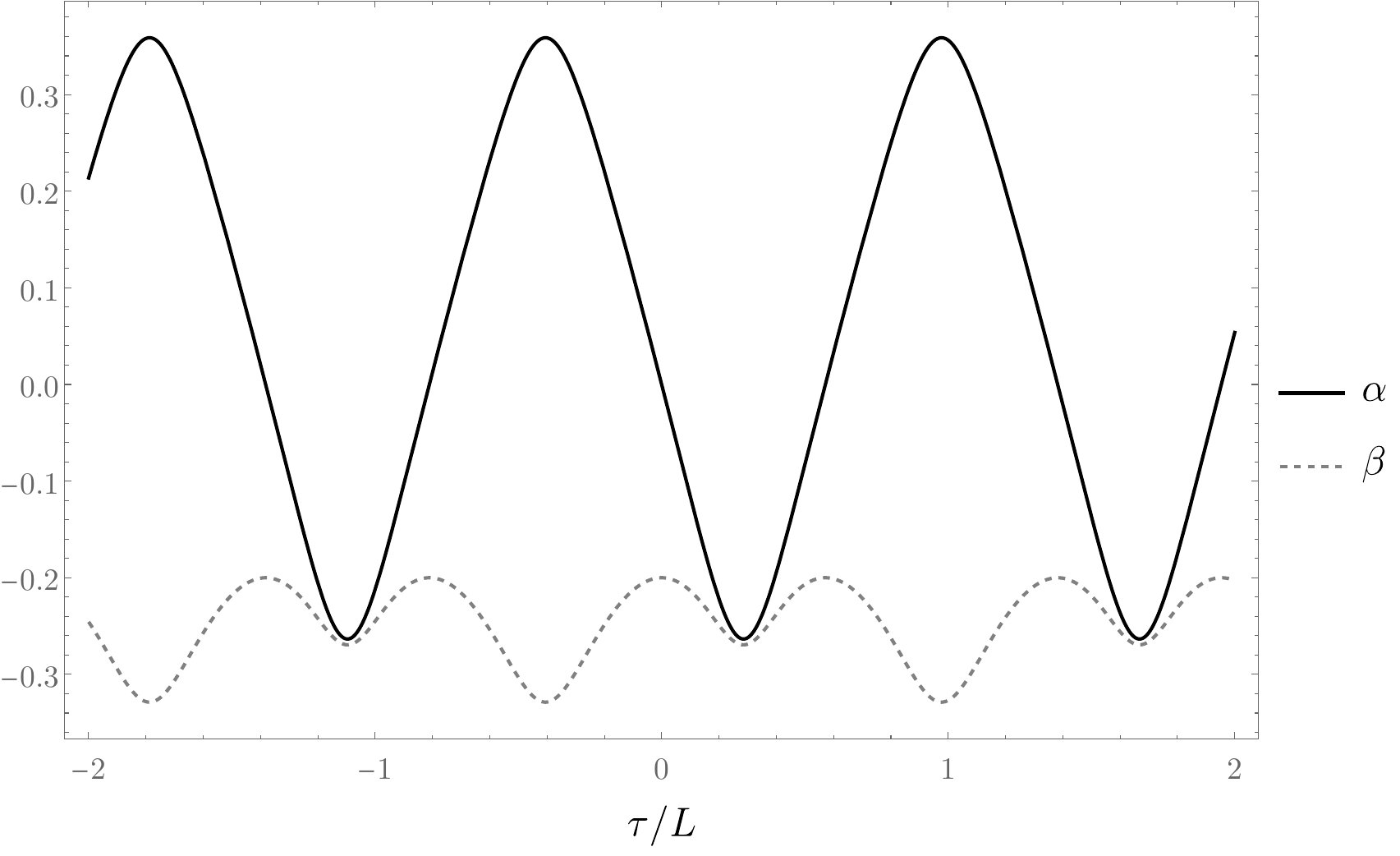}
    \caption{Time evolution of the moduli $\alpha$ (solid black line) and $\beta$ (dashed gray line) as functions of $\tau/L$, corresponding to the trajectory shown in Fig.~\ref{fig:moduli}. The extrema of $\alpha$ match the analytical predictions.}
    \label{fig:trajectpar}
\end{figure}

In Fig.~\ref{fig:trajectpar}, we plot $\alpha$ (solid black line) and $\beta$ (dashed gray line) as functions of $\tau/L$, using the same parameter values as those used to generate 
Fig.~\ref{fig:moduli}. The analysis in the previous paragraph correctly predicts the extrema of $\alpha$. We have also evaluated the Euclidean action for this configuration and found $ S_E / L^2 \approx -6.3547(0)\,n$, where $n$ is the number of cycles considered in the geometry. As $n$ can be taken arbitrarily large, the action becomes arbitrarily negative, as anticipated.

Here we have chosen a particularly simple case where the solution simply oscillates periodically in $x$, and where our restriction to $\gamma$ transforms the constraint into a relation that looks like energy conservation for a non-relativistic particle in a time-independent potential.    However, for future reference we note that we could also choose the curve $\gamma$ to change after each bounce off any turning point, so long as it still meets $V=0$ at the location of the desired bounce so that the solution remains continuous.  Our problem would then be formally equivalent to energy conservation in the presence of a time-dependent potential (that might change with each bounce).  This freedom will be useful at the beginning of the next section.  More general forms of time-dependence can also be introduced by choosing the relation between $\alpha$ and $\beta$ to depend explicitly on some time coordinate.

\begin{figure}[tb]
    \centering
    \includegraphics[width=0.6\textwidth]{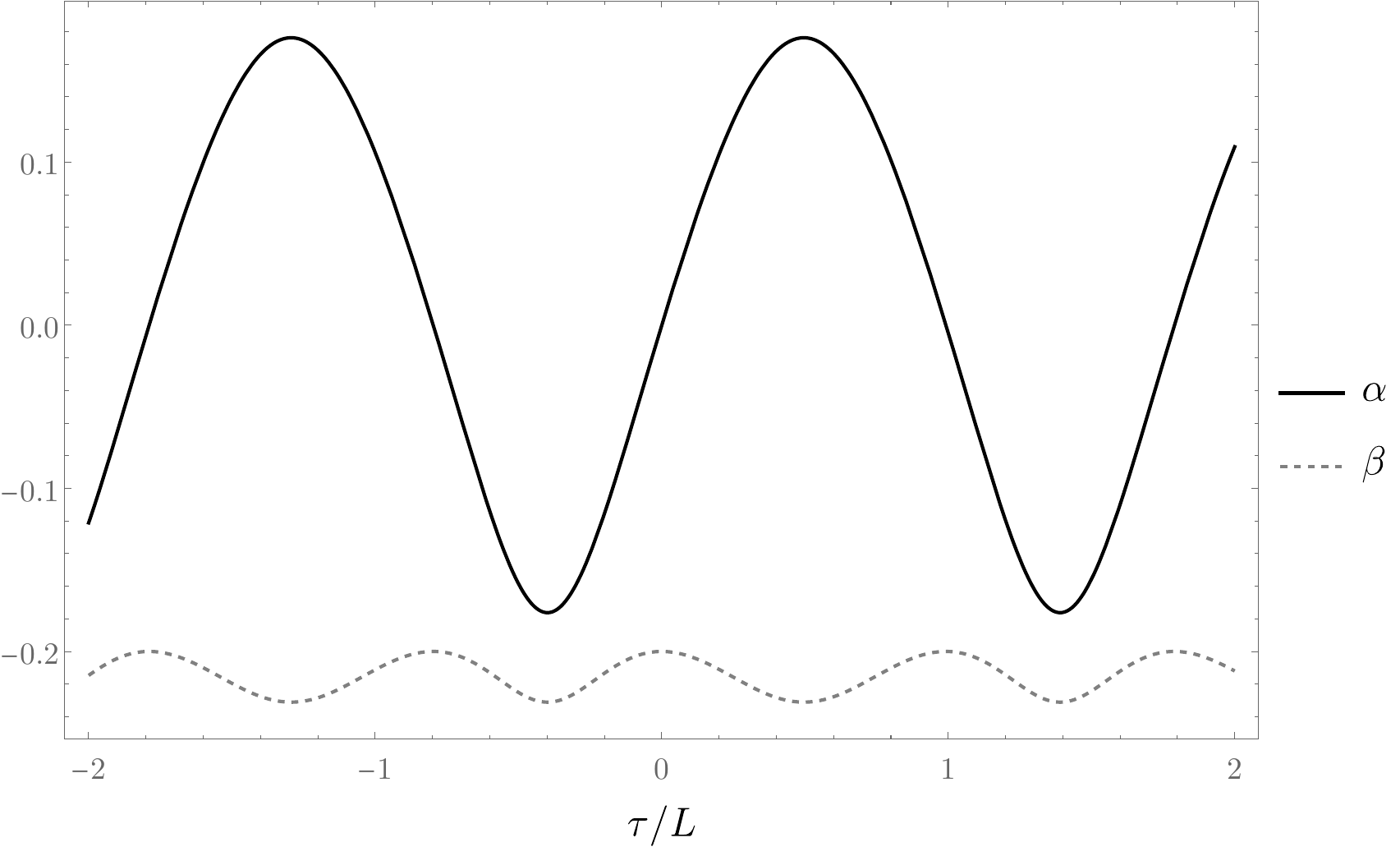}
    \caption{Time evolution of the moduli $\alpha$ (solid black line) and $\beta$ (dashed gray line) as functions of $\tau/L$, corresponding to the trajectory given in Eq.~(\ref{eq:traalphabeta}) with $\eta = -0.2$. Here, we set $\kappa = 0$.}
    \label{fig:asympF}
\end{figure}

The case with $\kappa = 0$ can be treated in a similar manner. We continue to use the trajectory for $\beta$ as given in Eq.~(\ref{eq:traalphabeta}). In this case, to ensure that $V < 0$, we simply require $\beta > \beta_0$. As before, we restrict to $|\alpha| < 1/2$. In Fig.~\ref{fig:asympF}, we plot $\alpha$ and $\beta$ as functions of $\tau/L$ for $\kappa = 0$, and observe the expected oscillatory behavior. The on-shell action can also be computed, yielding $ S_E/L^2 \approx -1.5453(4)\,n$, where $n$ denotes the number of oscillations. As in previous cases, taking $n$ to be large renders the Euclidean action unbounded from below. Note that for $\kappa = 0$, the parameter $L$ should be interpreted as an arbitrary length scale.

\section{Asymptotically Euclidean and thermal asymptotically flat examples}
\label{sec:asymflat}

The last example in 
Section~\ref{sec:nonposLambda} can be modified to give an asymptotically Euclidean metric with arbitrarily negative action which satisfies the constraints on a nested family of $S^3$'s. To do so, we merely set $\kappa=0$ and constrain the dynamics to an appropriate curve $\gamma$.  Before choosing $\gamma,$ let us note that the usual $S^3$ foliation of flat Euclidean space maintains isotropy on each slice and thus corresponds to motion along the curve $\gamma_0$ defined by setting $B=0$.  In particular, the metric on our Euclidean spacetime remains smooth when the spheres pinch off by taking $\alpha \rightarrow -\infty$ along $\gamma_0$.  

But we can now make the action arbitrarily negative by modifying the curve so that it no longer exactly coincides with $\gamma_0$ and, in particular, by 
making use of the freedom noted above to change $\gamma$ after each bounce.  Before the first bounce occurs, we choose any timelike curve in the $(\beta, \alpha)$ plane that starts at $B=0=\beta$ when $\alpha=+\infty$, but which reaches $V=0$  (i.e., $\beta_0=-\frac{1}{3}\ln 2$) at some finite value $\alpha=\alpha_-$ which defines the first bounce (we will use the term `bounce' for any encounter with a turning point, whether it is $\alpha_-$ or $\alpha_+)$.  After this first bounce, and until bounce $2n$ (for some positive integer $n$), we then choose $\gamma$ to be of the form discussed in the previous section so that the $\alpha$ executes $n$ oscillations between $\alpha_-$ and some $\alpha_+ > \alpha_-$.  By taking $n$ large, we can make the action arbitrarily negative.  After bounce $2n$ we then simply choose $\gamma$ to be any timelike curve from  $(\alpha_+, \beta_+)$  that coincides with $\gamma_0$ for  sufficiently negative $\alpha$ (say, for $\alpha < \alpha_+ + 2\beta_0)$.  This guarantees that the spacetime metric will remain smooth when the spheres pinch off and that the evolution before bounce $1$ and after bounce $2n$ make only finite contributions to the action.  Thus, as before, the action can be made arbitrarily negative by taking $n$ to be large.  

Now, one might ask if the above might somehow be an artifact of taking the foliation to pinch off (so that the last leaf in the foliation is the degenerate metric with $a=0$). However, if one considers spacetimes with two disjoint asymptotically Euclidean boundaries (i.e., spacetime wormholes), it is straightforward to construct examples with nondegenerate foliations by instead choosing the curve $\gamma$ after bounce $2n-1$ to again follow a timelike curve from $(\alpha_-, \beta_-)$ that coincides with $\gamma_0$ at sufficiently large positive values of $\alpha$.

One may also construct examples using foliations in which each slice is noncompact and in fact is asymptotically Euclidean.  In particular, we can construct examples which are asymptotic to thermal flat space. This is easiest to achieve if we add a massless complex scalar field $\Phi$. The bulk part of the action then becomes $S_E = -\frac{1}{16\pi} \int (R - |\nabla \Phi |^2 )\,\sqrt{g}\,{\rm d}^4x$.
The Hamiltonian constraint  is now
\begin{equation}\label{eq:asymflat}
{\cal R} + K_{ab} K^{ab} - K^2 = -2 G_{ab}n^a n^b = |D_a \Phi|^2 - |\dot \Phi|^2,
\end{equation}
where $n^a$ is a unit normal to our surface and $D_a$ is the spatial covariant derivative.  The advantage of working with a complex field is that we can add a time-dependent phase and keep the metric static.

 Consider then the metric 
 \begin{equation}
 \label{eq:sphmass}
 {\rm d}s^2 = {\rm d} \tau^2 + \frac{{\rm d}r^2}{1 - \frac{2m(r)}{r}} + r^2 {\rm d}\Omega^2.
 \end{equation}
 We choose our foliation to be the static slices of constant $\tau $. These surfaces clearly have $K_{ab} = 0$. We will construct an example which is periodic in $\tau$, appropriate for path integrals with thermal boundary conditions. Pick a period $\beta$ for $\tau$ and set $\Phi = e^{i\omega \tau} \phi(r)$ with $\omega = 2\pi n/\beta$. 
Choose $\phi(r)$ to be a small constant  $\epsilon$ inside a ball of radius $r_0$. Then the $|\dot \Phi|^2 = \epsilon^2 \omega^2$ term in the constraint \eqref{eq:asymflat} acts like a constant negative energy density. We want to take $\omega$ very large, so $\epsilon^2 \omega^2 \gg 1$. As is well known, for metrics of the form \eqref{eq:sphmass} with a regular origin the Hamiltonian constraint equates $m(r)$ with the integral of this effective energy density.  For $r \le r_0$ we thus find 
\be\label{eq:mass}
m(r) = - 2\pi \int |\dot \Phi|^2 r^2 {\rm d}r =  -\frac{2\pi}{3} \epsilon^2 \omega^2 r^3.
\ee

Outside the ball, we can take $\phi(r)$ to vanish for $r > r_0 + \epsilon$.  
In the transition region, $r_0 < r < r_0+\epsilon$, we have $\partial_r \phi \sim 1$, and $g^{rr}$ is dominated by the mass term, so $|D_a\Phi|^2$ acts like a large positive energy density. 
 Fortunately, this is integrated over a volume which is proportional to $\epsilon$.  It's contribution thus  remains small compared to the negative contribution inside the ball. As a result, the total mass is 
\be
M \sim - \frac{2\pi}{3}\epsilon^2 \omega^2 r_0^3.
\ee

Writing the action in Hamiltonian form and imposing the constraint we find
\be 
S_E = \frac{1}{8\pi} \int \left(K_{ab} K^{ab} - K^2  +|\dot \Phi|^2 \right) \sqrt{h}\ {\rm d}\tau\,{\rm d}^3x +  \int M {\rm d}\tau,
\ee
where $h_{ab}$ is the spatial metric. Using (\ref{eq:mass}) and $K_{ab} = 0$, for $r< r_0$ the first term  is
\be
\frac{1}{2} \int   \frac{\epsilon^2 \omega^2 r^2}{\sqrt{1+ c\epsilon^2 \omega^2 r^2}} {\rm d}r {\rm d}\tau
\ee
where $c = 4\pi/3$. So for large $\epsilon^2 \omega^2$, the integral over the ball $r\le r_0$ is $\mathcal{O}(\epsilon \omega r_0^2\beta)$.  It thus makes a positive contibution, but one that is  smaller in magnitude than the (negative) mass term.  Since $\dot{\Phi}$ vanishes for $r> r_0+\epsilon,$ it remains only to include the transition region $r_0 < r < r_0 + \epsilon$.  This changes the integral  only  by a subleading amount which is smaller by a factor of $\epsilon/r_0$.

The net result is thus that the action is dominated by the mass term
\be 
S_E\sim - \epsilon^2 \omega^2 r_0^3 \beta
\ee
which can be made arbitrarily negative.
 A similar construction  also works in {other dimensions $D \ge 3$.}  For $D=5$ one may also use a Bianchi IX-like construction involving squashed spheres which allows one to perform an analogous argument using vacuum Einstein-Hilbert gravity without the addition of a complex scalar.  This latter version will be illustrated in the construction of asymptotically AdS examples below.

\section{Asymptotically {thermal} AdS examples}
\label{sec:AdS}

When $\Lambda < 0$ one can again modify the example of Section \ref{sec:nonposLambda} 
 to give an asymptotically AdS metric with arbitrarily negative action which satisfies the constraints on a nested family of $S^3$'s.   As for the $\Lambda=0$ construction described at the beginning of section \ref{sec:asymflat},   one may proceed by first choosing a value of the parameter $\eta$ which makes the curve \eqref{eq:traalphabeta} timelike in the region of negative $V$ and which intersects $V=0$ at two points $(\alpha_+,\beta_+)$ and $(\alpha_-,\beta_-)$.  One then simply first locates the curve $\gamma_0$ that corresponds to the spherical slicing of Euclidean AdS and then designs a timelike curve $\gamma$ that follows $\gamma_0$ in from the region of large positive $\alpha$ but which reaches the $V=0$ surface at $\alpha_-$.  After the first bounce at $\alpha_-$,  one takes $\gamma$ to be given by \eqref{eq:traalphabeta} until the completion of bounce $2n$ (which occurs at $\alpha_+$). After this point, one chooses any timelike curve from $(\alpha_-, \beta_-)$ that coincides with $\gamma_0$ for sufficiently negative values of $\alpha$.  Taking $n$  large then gives smooth spacetimes that satisfy the constraints on each $S^3$ but with arbitrarily negative actions. 

Once again, however, we would also like to provide an example with noncompact slices where each slice satisfies standard AdS asymptotics.  In analogy with section \ref{sec:asymflat}, 
we now give an example which is asymptotic to thermal AdS. One way to do this is to include a massless complex scalar field and repeat the construction of the previous section. One finds that adding $\Lambda < 0$ causes only minor changes, and does not prevent the action from becoming arbitrarily negative.

In five spacetime dimensions it is also straightforward to numerically find pure vacuum examples asymptotic to thermal AdS.  Using the $\sigma_i$ defined in \eqref{eq:sigmas}, let us define the following 1$-$forms
\begin{equation}
\sigma_{\pm}=\sigma_1\pm i \sigma_2\,.
\end{equation}
Consider now the line element
\begin{multline}
{\rm d}s^2=\frac{f(r)}{H(r)}{\rm d}\tau^2+\frac{{\rm d}r^2}{f(r)}
+\frac{r^2}{4}\Big[\sqrt{1+\left|\Psi(\tau,r)\right|^2}\sigma_+ \sigma_-
\\
+\frac{\bar{\Psi}(\tau,r)}{2} \sigma_+^2+\frac{\Psi(\tau,r)}{2} \sigma_-^2+H(r)\left(\sigma_3+2 \Omega(r){\rm d}\tau\right)^2\Big]\,,
\end{multline}
with
\begin{equation}
\Psi(\tau,r)=e^{\frac{2 i n \pi  \tau }{\beta }}\Phi(r)
\label{eq:defP}
\end{equation}
where $\beta$ is the length of the thermal circle, i.e. $\tau\sim\tau+\beta$, and $n\in\mathbb{Z}$. In particular, we take the metric to have an exact symmetry under the simultaneous action of a translation in $\tau$ and a rotation of $\sigma_1$ into $\sigma_2$ (i.e., a shift of the angle $\psi$ in \eqref{eq:sigmas}). We are interested in finding solutions that asymptote to the thermal Einstein static cylinder
\begin{equation}\label{eq:staticcyl}
{\rm d}s^2_{\partial}={\rm d}\tau^2+\frac{L^2}{4}(\sigma_+ \sigma_- +\sigma^2_3)={\rm d}\tau^2+L^2 {\rm d}\Omega_3^2\,,
\end{equation}
where ${\rm d}\Omega_3^2$ is the line element of a round three-sphere of unit radius. Furthermore, since we want to use the usual counter-term expansion, we want deviations from the thermal Einstein static cylinder to start at order $\mathcal{O}(z^2)$, where $z$ is the standard Fefferman-Graham coordinate with $z=0$ marking the location of the conformal boundary.

Let us define
\begin{equation}
f(r)=\frac{r^2}{L^2}+H(r)-\frac{2m(r)}{r^2}\, ,
\end{equation}
so that the Hamiltonian and momentum constraints take the form
\begin{subequations}
\begin{multline}
\mathcal{H}\equiv m'(r)+\frac{m'(r) r}{6}\frac{H'(r)}{H(r)}+\frac{2 r^4-8 L^2 m(r)}{6 L^2}\frac{H^\prime(r)}{H(r)}-\frac{1}{12}\frac{r^3 f(r) \Phi '(r)^2}{1+\Phi (r)^2}
\\
+\frac{1}{12} r^5 H(r)^2 \Omega
   '(r)^2-\frac{f(r)\left[r^7 H^\prime(r)\right]^\prime}{6r^4 H(r)}+\frac{r^4-2 L^2 m(r)}{12 L^2
   H(r)^2} r H'(r)^2
   \\
   +\frac{4r}{3}\sqrt{1+\Phi (r)^2}-\frac{4 r}{3 H(r)}\left[H(r)^2+\Phi (r)^2\right]+\frac{r^3 H(r) \Phi (r)^2 }{3 \beta ^2 f(r)}\left[n \pi +2 \beta  \Omega (r)\right]^2=0
\end{multline}
and
\begin{equation}
\mathcal{P}\equiv\left[H(r)^2 r^5 \Omega^\prime(r)\right]^{\prime}-\frac{8 L^2 r^3 H(r) }{\beta  f(r) L^2} \Phi (r)^2\left[n \pi +2 \beta  \Omega (r)\right]=0
\end{equation}
\end{subequations}
respectively.

For the action, we take
\begin{multline}
S_E=-\frac{1}{16\pi }\int_{\mathcal{M}}{\rm d}^5 x\sqrt{g}\left(R+\frac{12}{L^2}\right)-\frac{1}{8 \pi }\int_{\partial \mathcal{M}}{\rm d}^4x\sqrt{h}\tilde{K}
\\
+\frac{3}{8\pi L }\int_{\partial \mathcal{M}}{\rm d}^4x\sqrt{h}+\frac{L}{32\pi }\int_{\partial \mathcal{M}}{\rm d}^4x\sqrt{h}R_h\,.
\end{multline}
where the last two terms are the usual counter-terms in AdS$_5$. We have checked that the above renders the action finite, so long as
\begin{multline}
m(r)=L^2\widetilde{m}_0+\ldots\,,\quad H(r)=1+\widetilde{H}_0\frac{L^4}{r^4}+\ldots\,,
\\
\Omega(r)=\widetilde{\Omega}_0\frac{L^4}{r^4}+\ldots\quad\text{and}\quad \Phi=\widetilde{\Phi}_0\frac{L^4}{r^4}+\ldots
\end{multline}
as $r\to+\infty$. In the expression above, the $\ldots$ denote subleading terms in $r$. In particular, the above results hold for on-shell solutions of the Einstein equation endowed with a negative cosmological constant.

Evaluating the action, and using the constraints,  we find
\begin{equation}
S_E=\frac{n\,\pi^2}{2\beta 
}\int_0^{+\infty}\frac{r^3 H(r) \left[n \pi +2 \beta  \Omega (r)\right] \Phi (r)^2}{f(r)}{\rm d}r+\frac{L^2 \pi  \beta}{32 }  \left(3+16 \widetilde{H}_0+24 \widetilde{m}_0\right)\,,
\label{eq:action}
\end{equation}
where the first term is manifestly finite for our boundary conditions. The second term is simply $\beta E$, with $E$ the energy as computed using the standard holographic renormalisation procedure. In detail, we can introduce Fefferman-Graham coordinates, via
\begin{equation}
r=\frac{L^2}{z} \left(1-\frac{1}{4}\frac{z^2}{L^2}+\frac{\tilde{m}_0}{4}\frac{z^4}{ L^4}+\ldots\right)
\end{equation}
which brings the metric ansatz to the form
\begin{subequations}
\begin{equation}
{\rm d}s^2=\frac{L^2}{z^2}\left({\rm d}z^2+{\rm d}s^2_{\partial}+z^2 {\rm d}s^2_2+z^4{\rm d}s^2_4+\ldots\right),
\end{equation}
where
\begin{equation}
{\rm d}s^2_2=\frac{1}{2 L^2}{\rm d}\tau^2-\frac{1}{8} \left(\sigma _+ \sigma _-+\sigma _3^2\right)
\end{equation}
and
\begin{multline}
{\rm d}s^2_4=\frac{1-16 \tilde{H}_0-24 \tilde{m}_0}{16 L^4}{\rm d}\tau^2 +\frac{1+8 \tilde{m}_0}{64 L^2}\sigma _- \sigma _++\frac{1+16 \tilde{H}_0+8
   \tilde{m}_0}{64 L^2}\sigma _3^2
   \\
   +e^{\frac{2 i n \pi  \tau }{\beta }} \frac{\tilde{\Phi }_0}{8 L^2} \sigma _-^2+e^{-\frac{2 i n \pi  \tau }{\beta }} \frac{\tilde{\Phi }_0}{8
   L^2} \sigma _+^2+\frac{\tilde{\Omega }_0}{L^2}{\rm d}\tau\,\sigma _3.
\end{multline}
\end{subequations}
Using standard results in holographic renormalisation \cite{Bianchi:2001de,Bianchi:2001kw}, one can find
\begin{multline}
T^{{\rm 4D}}_{\mu \nu}\,{\rm d}x^{\mu}{\rm d}x^{\nu}=\frac{L}{64 \pi  
} \Bigg(-\frac{3+16 \tilde{H}_0+24 \tilde{m}_0}{L^2} {\rm d}\tau^2+\frac{1+8 \tilde{m}_0}{4}\sigma _+ \sigma _-
\\
+2 e^{-\frac{2 i n \pi  \tau }{\beta }} \tilde{\Phi
   }_0 \sigma _+^2+2 e^{\frac{2 i n \pi  \tau }{\beta }} \tilde{\Phi }_0 \sigma _-^2+\frac{1+16 \tilde{H}_0+8 \tilde{m}_0}{4}\sigma _3^2+16 \tilde{\Omega }_0 \text{d$\tau $} \sigma
   _3\Bigg),
\end{multline}
confirming that the last term in Eq.~(\ref{eq:action}) is indeed the energy multiplied by $\beta$.

We are interested in smooth geometries, with a smooth center at $r=0$. Near $r=0$ this requires
\begin{subequations}
\begin{equation}
m(r)=\mathcal{O}(r^4)\,, H(r)=1+\mathcal{O}(r^2)\,,\Phi(r)=\mathcal{O}(r^2)\quad\text{and}\quad \Omega(r)=\widehat{\Omega}_0+\mathcal{O}(r^2)\,,
\end{equation}
with $\widehat{\Omega}_0$ a constant. In order to impose these conditions, together with the required asymptotic behaviour, we perform a change of variables of the following form:
\begin{align}\label{eq:defS}
&\Phi(r)=\frac{r^2}{L^2}\frac{1}{\displaystyle\left(1+\frac{r^2}{L^2}\right)^{3}}S_1(r)\,,
\\
&H(r)=1+\frac{r^2}{L^2}\frac{1}{\displaystyle\left(1+\frac{r^2}{L^2}\right)^{3}}S_2(r)\,,
\\
&m(r)=\frac{r^4}{L^2}\frac{1}{\displaystyle\left(1+\frac{r^2}{L^2}\right)^{2}}Q_1(r)\,,
\\
&\Omega(r)=\frac{1}{\displaystyle\left(1+\frac{r^2}{L^2}\right)^{2}}Q_2(r)\,.
\end{align}
\end{subequations}

We will pick a profile for $S_1$ and $S_2$, and use the Hamiltonian and momentum constraints to solve for $Q_1$ and $Q_2$, respectively. In the numerics, it is convenient to use a compact coordinate, which we will take to be
\begin{equation}
y=\frac{1}{\displaystyle 1+\frac{r^2}{L^2}}\,,
\end{equation}
with the origin being $y=1$ and asymptotic infinity being $y=0$. Note that $\widetilde{m}_0=q_1(0)$ and $\widetilde{H}_0=s_2(0)$, where we defined
\begin{equation}
q_i(y)\equiv Q_i\left(\frac{L\sqrt{1-y}}{y}\right)\quad\text{and}\quad s_i(y)\equiv S_i\left(\frac{L\sqrt{1-y}}{y}\right)\,.
\end{equation}
The boundary conditions in terms of the $q_i(y)$ are easily determined from those of the $Q_i(r)$. In order to evaluate the action, we first change to the $q_i$ and $s_i$ variables and have to also take into account the change of variable. Once the dust settles, we find
\begin{subequations}
\begin{multline}
\frac{ S_E}{L^3}=\int_0^1\frac{n \pi ^2 (1-y)^3 y^2  s_1(y)^2h_1(y) h_2(y) h_3(y)}{4 \tilde{\beta } \left[1-2 (1-y) y^2q_1(y)+(1-y) y^3 s_2(y)\right]}{\rm d}y
\\
+\frac{1}{32} \pi  \tilde{\beta } \left[3+24 q_1(0)+16 S_2(0)\right]\,,
\end{multline}
where $\tilde{\beta}\equiv \beta/L$ and
\begin{align}
&h_1(y)\equiv n \pi +2 y^2 \tilde{\beta } q_2(y)\,,\nonumber
\\
&h_2(y)\equiv 2+(1-y)^2 y^4 s_1(y)^2\,,
\\
&h_3(y)\equiv 1+(1-y) y^2 s_2(y)\,.\nonumber
\end{align}
\end{subequations}

\begin{figure}[tb]
    \centering
    \includegraphics[width=0.6\textwidth]{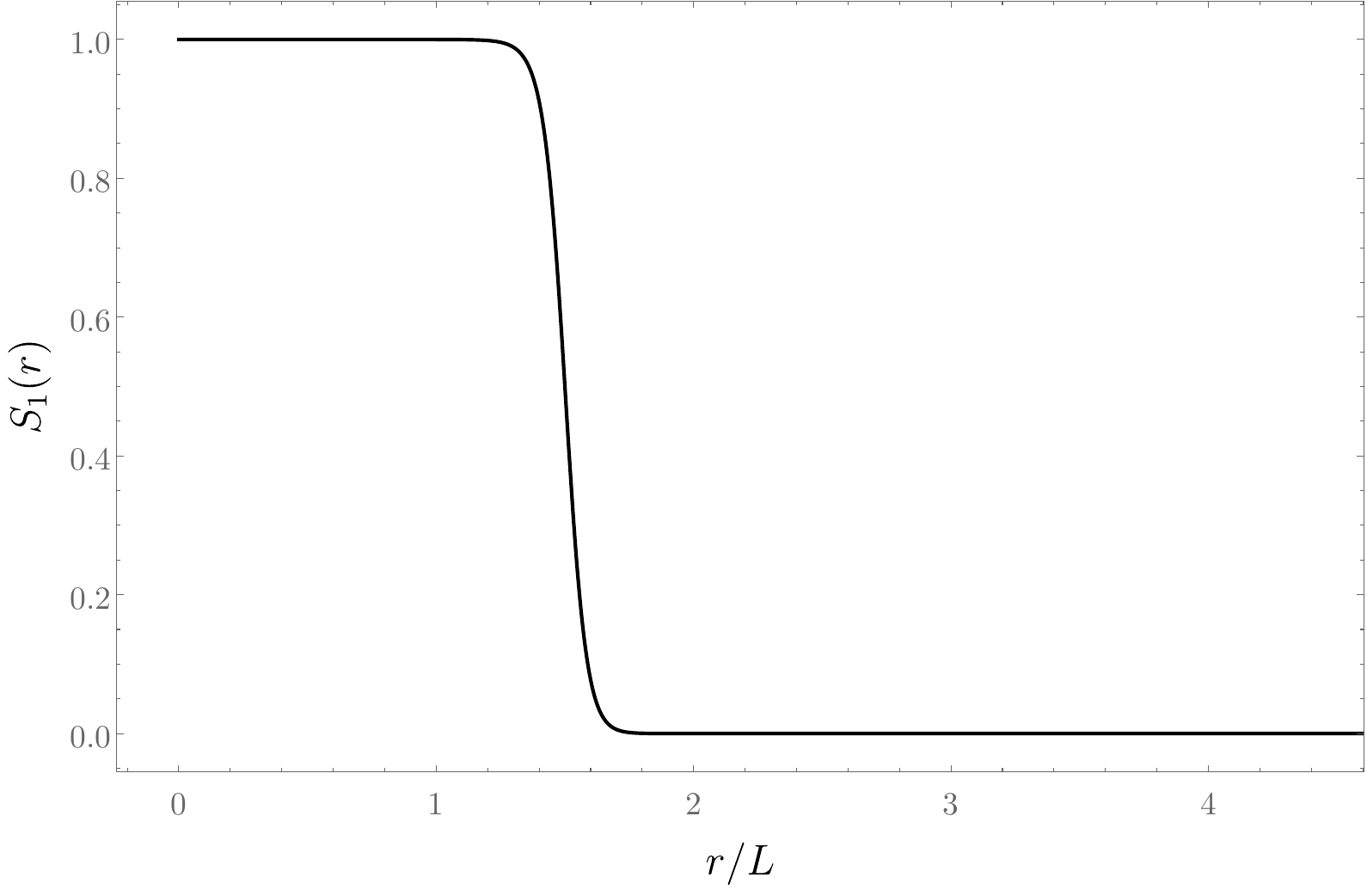}
    \caption{$S_1(r)$ (defined in \eqref{eq:defP} and \eqref{eq:defS}) as a function of $r/L$ for the profile given in Eq.~(\ref{eq:prof}) with  $A_0=1$, and for $y_0=3/2$ and $\tilde{\sigma}=1/2$.}
    \label{fig:profile}
\end{figure}

For our numerical experiments, we take
\begin{subequations}
\begin{align}
&s_1(y)=\frac{ A_0}{2} \left\{1-\tanh \left[\frac{1}{\tilde{\sigma }^2}\left(\frac{1-y}{y}-y_0^2\right)\right]\right\}\label{eq:prof}
\\
&s_2(y)=B_0
\end{align}
\end{subequations}%
where $A_0$, $B_0$, $y_0$ and $\tilde{\sigma}$ are constants that we can dial. We wish the profile to be almost constant near a region $r\leq r_0$, and to drop to zero at large $r$. We achieve this by setting $y_0=3/2$ and $\tilde{\sigma}=1/2$, which we will fix henceforth. In Fig.~\ref{fig:profile} we plot the profile $S_1(r)$ for $A_0=1$, and for $y_0=3/2$ and $\tilde{\sigma}=1/2$.

\begin{figure}[tb]
    \centering
    \includegraphics[width=0.8\textwidth]{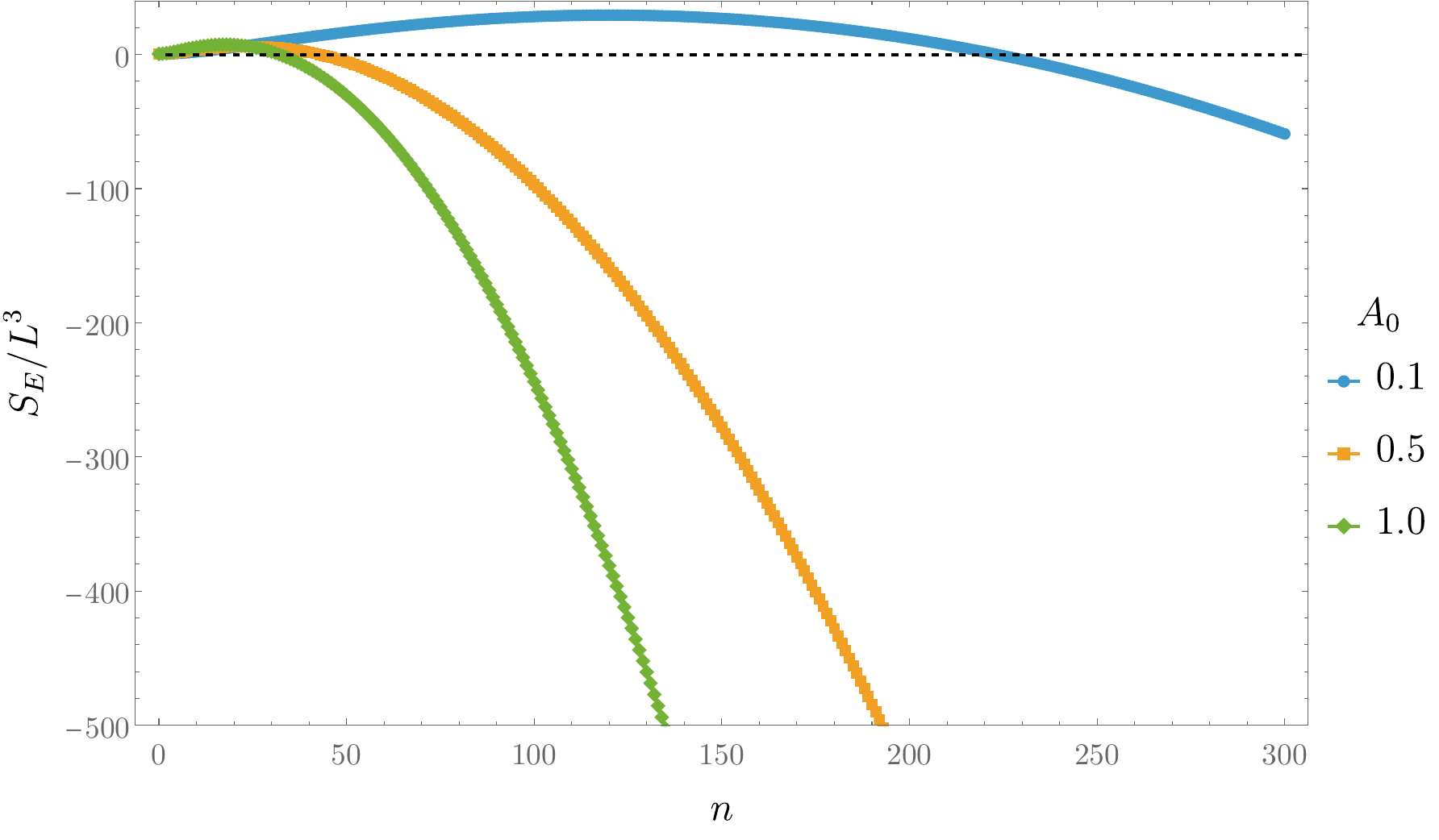}
    \caption{$ S_E/L^3$ as a function of $n$, for $B_0=\sqrt{1+n^2}$, $y_0=3/2$ and $\tilde{\sigma}=1/2$ and $\tilde{\beta}=1$. Different symbols represent different fixed values of $A_0$ as labeled on the right.}
    \label{fig:example}
\end{figure}

We find that if we take $B_0\sim |n|$ at large $n$, then, for any fixed values of $A_0$, we can find a value $n_\star$ such the for $|n|>n_{\star}$ the action becomes increasingly negative. In the example below we take $B_0=\sqrt{1+n^2}$. The smaller the value of $A_0$, the larger the value of $n$ that does the job. In Fig.~\ref{fig:example} we plot $ S_E/L^3$ as a function of $n$, for fixed $\tilde{\beta}=1$ and for the values of $A_0$ labeled on the right.

\begin{figure}[tb]
    \centering
    \includegraphics[width=0.53\textwidth]{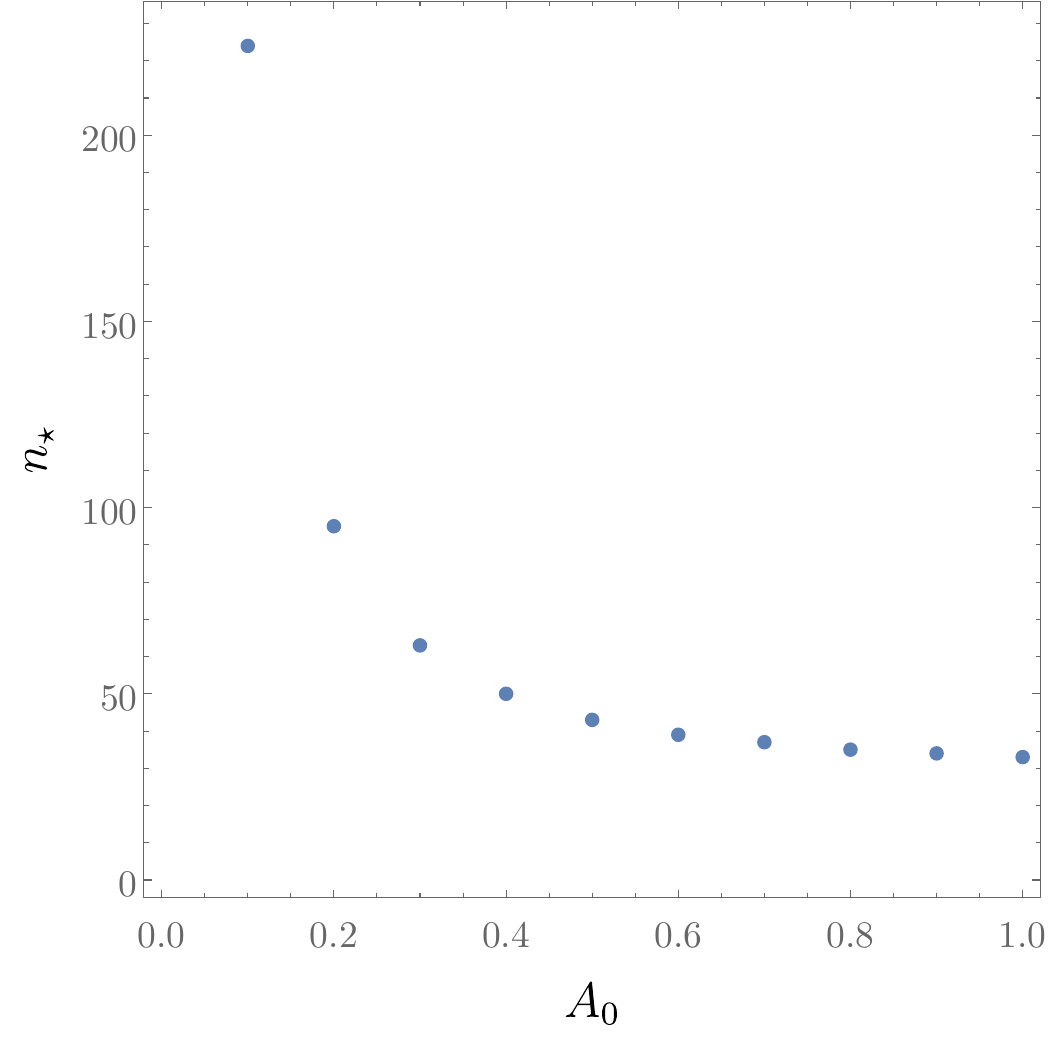}
    \caption{$n_{\star}$, the critical value of $n$ above which $S_E$ becomes negative, as a function of $A_0$ for fixed $y_0=3/2$ and $\tilde{\sigma}=1/2$ and $\tilde{\beta}=1$.}
    \label{fig:acrit}
\end{figure}

In Fig.~\ref{fig:acrit} we plot $n_{\star}$ as a function of $A_0$ for fixed values of $B_0=\sqrt{1+n^2}$, $y_0=3/2$ and $\tilde{\sigma}=1/2$ and $\tilde{\beta}=1$. As indicated above, the smaller the values of $A_0$, the larger values of $n$ we need to reach in order for the action to become arbitrarily negative.

 %\subsection{Asymptotically $AdS_4$ example}

%In order to convert the above example to an asymptotically (globally) $AdS_4$ example, we proceed as follows. We take the above construction for $|\tau| < T_0$ where $T_0$ is a long time, producing a configuration with action $I\sim - \epsilon^2 \omega^2 r_0^3 T_0$. We then let $\phi$ become time dependent in a transition region $T_0 < |\tau| < T_1$, so that $\phi = 0$ for $|\tau| > T_1$. Since $\phi = O(\epsilon)$, for  $|\tau| < T_0$, it does not have to change much to vanish. Furthermore, since $T_1$ is arbitrary, we can make $\partial_\tau \phi$ as small as we want. (So we can treat it perturbatively if needed.) We just have to show that we can solve the constraints in this transition region. It doesn't matter what the action is for this transition region since we can always make $T_0$ arbitrarily large.

%One way to solve the constraints might be the following. We can write the spatial metric as a conformal factor times a fixed (unit) hyperbolic metric. Inside the ball of radius $r_0$, the problem is then just an open FRW universe. We can choose a time dependent $\phi(\tau)$ and solve for the scale factor. Outside the ball, the equations are more complicated since $ \phi = \phi(r,\tau)$ and the momentum constraint is now nontrivial. But it seems likely that if we make a general enough metric ansatz (perhaps including  a radial shift vector), we should be able to solve them.

\section{Rescaling to constant scalar curvature}
\label{sec:constR}
 As mentioned in the introduction, an early proposal for dealing with the unbounded nature of the action was to try to locally rescale (i.e., to Weyl transform) asymptotically Euclidean metrics to have zero scalar curvature. Unfortunately, this is not possible in all cases.  

In this section we consider this proposal with different boundary conditions.  As in the original proposal, the idea will be to integrate over real Euclidean metrics with a given constant scalar curvature $R$ and then to separately integrate the Weyl factor over some contour that allows that integral to converge.  Below, we study convergence of the integral over metrics with scalar curvature $R$.

Before turning to the asymptotically AdS case in section \ref{sec:thermalR}, let us briefly discuss the case of compact Riemannian manifolds without boundary. The question of whether one can always rescale a metric on such a manifold to have constant scalar curvature $R$ is known as the Yamabe problem. After considerable effort by a number of mathematicians, Schoen \cite{SCH} proved that the answer is yes. The sign of $R$ is determined by the sign of the lowest eigenvalue of the conformally invariant scalar Laplacian.  

One can find geometries of arbitrarily large spacetime volume $V$ for either sign of $R$.  In particular, 
fixed $R>0$ again allows the construction of chains of spheres  connected by thin necks as in Fig. \ref{fig:illu},   while there are compact hyperbolic manifolds with arbitrarily large volume for fixed $R<0$ (and $R=0$ clearly allows large tori).  Since the Euclidean action is proportional to $(2\Lambda -R)V$, the action is bounded below precisely when we choose $R\le2 \Lambda$.  For $\Lambda>0$ we can certainly make such a choice, but for $\Lambda <0$ the action will be unbounded below for manifolds that require $R>0$ (e.g., with spherical topology).  As a simple example, one may consider the following four-dimensional metric
\begin{equation}
{\rm d}s^2={\rm d}\tau^2+a(\tau)^2{\rm d\Omega_3^2}\,.
\end{equation}
Imposing
\begin{equation}
R=\frac{12}{L^2}
\end{equation}
yields
\begin{equation}
a(\tau)=\frac{L}{\sqrt{2}} \sqrt{1-C \cos \left[\frac{2 (\tau-\tau_0) }{L}\right]}\,,
\end{equation}
where $C$ and $\tau_0$ are constants of integration. For $-1<C<1$, the above solution represents a periodic configuration with period $\Delta \tau=\pi L$. 

If one were content to restrict to manifolds that allow $R<0$, then for $\Lambda <0$ the action is bounded below if we fix $R$ so that we have $|R| \ge 2|\Lambda|$.  In particular, the action is bounded below for $R = \frac{2D\Lambda}{D-2}$, as is the case in the empty AdS space that would solve the Einstein equations with the given value of $\Lambda$.
Suppose, however, that instead of a cosmological constant there is a potential $V(\phi)$ of some scalar field $\phi.$  We then see that the action with any fixed $|R|$ remains unbounded below when $V(\phi)$ takes sufficiently negative values; e.g., if one allows scalar fields with negative mass (squared)\footnote{\label{foot:V} It is possible that this problem could be avoided by instead choosing the Weyl rescaling to set $R$ proportional to $V(\phi)$ with an appropriate coefficient.  However, we will not further explore this idea here.}.  We will see below that such fields also lead to arbitrarily negative fixed-$R$ actions for spacetimes asymptotic to thermal AdS.

\subsection{Thermal AdS$_4$ boundary conditions}
\label{sec:thermalR}

In AdS, there are matter fields with negative potentials which still satisfy a positive energy theorem. In fact, many such fields arise in supergravity and play important roles in holography. In this section we include one of these fields and show that the action can become arbitrarily negative keeping the scalar curvature fixed to a negative constant.\footnote{We thank Neta Engelhardt for suggesting that we include this matter field. It will be important that this field has a potential that falls off much faster than a $\mu^2<0$ term.}

We consider the following action:
\begin{subequations}
\begin{multline}
S_E = -\frac{1}{16\pi } \int_{\mathcal{M}} {\rm d}^4x \sqrt{g} \left[ R - \frac{1}{2} \nabla_a \Phi \nabla^a \Phi - V(\Phi) \right] - \frac{1}{8\pi } \int_{\partial \mathcal{M}} {\rm d}^3x \sqrt{h}\,K \\
+ \frac{1}{4\pi L}  \int_{\partial \mathcal{M}} {\rm d}^3x \sqrt{h} + \frac{L}{16\pi }  \int_{\partial \mathcal{M}} {\rm d}^3x \sqrt{h} R_h + \frac{1}{32\pi L } \int_{\partial \mathcal{M}} {\rm d}^3x \sqrt{h} \Phi^2\,,
\label{eq:scalarsugra}
\end{multline}
where
\begin{equation}
V(\Phi) = -\frac{2}{L^2} \left(2 + \cosh \Phi \right)\,.
\end{equation}
\end{subequations}
This can be obtained via a consistent truncation of eleven-dimensional supergravity  \cite{Cvetic:1999xp} \footnote{When the metric is positive definite, the four-form of eleven-dimensional supergravity becomes imaginary, just like an electric field.}.

The last three terms in Eq.~(\ref{eq:scalarsugra}) correspond to the standard counterterms in AdS$_4$ with a scalar field of mass $\mu^2 L^2 = -2$. The scalar field $\Phi$ admits two possible linear boundary conditions that preserve conformal invariance; in this section, we consider the case of standard boundary conditions, for which the last boundary term in Eq.~(\ref{eq:scalarsugra}) does not contribute.

Although the potential $V(\Phi)$ is negative - becoming exponentially negative at large $\Phi$ - solutions to the equations of motion derived from Eq.~(\ref{eq:scalarsugra}) obey a positive energy theorem \cite{Boucher:1984yx,Townsend:1984iu,Amsel:2007im}.  The origin of this theorem lies in the fact that $V$ can be expressed in terms of another function $W$, called the pre-potential. In fact, positive energy requires that $V$ can be recovered from two different prepotentials, $W^\pm$, 
 for all values of $\Phi$ \cite{Amsel:2007im}. In particular,\footnote{In \cite{Amsel:2007im}, the existence of $W^-$ was shown to guarantee the existence of $W^+$, but not vice versa.}
\begin{equation}
W^-(\Phi) = \frac{\sqrt{2}}{L} \cosh\left(\frac{\Phi}{2}\right)\,.
\end{equation}

We focus on configurations where the boundary metric is
\begin{equation}
{\rm d}s^2_{\partial} = {\rm d}\tau^2 + L^2 {\rm d}\Omega_2^2\,,
\end{equation}
where ${\rm d}\Omega_2^2$ is the line element of a round two-sphere of unit radius and $\tau \sim \tau + \beta$. For the bulk metric, we consider the following ansatz:
\begin{equation}
{\rm d}s^2 = f(r) {\rm d}\tau^2 + \frac{{\rm d}r^2}{f(r)} + r^2 {\rm d}\Omega_2^2\,.
\end{equation}
We assume that $f(r)$ has a bolt at $r = r_+$, so that $f(r_+) = 0$. Regularity at the bolt and the periodicity of the thermal circle then require
\begin{equation}
\beta = \frac{4\pi}{|f'(r_+)|}\,.
\end{equation}

Imposing a constant scalar curvature consistent with AdS asymptotics,
\begin{equation}
R + \frac{12}{L^2} = 0\,,
\end{equation}
we find that $f(r)$ must take the form
\begin{equation}
f(r) = \frac{r^2}{L^2} + 1 + \frac{2 r_+}{r} \left( \frac{2\pi r_+}{\beta} - 1 - \frac{2 r_+^2}{L^2} \right) + \frac{r_+^2}{r^2} \left( 1 + \frac{3 r_+^2}{L^2} - \frac{4\pi r_+}{\beta} \right)\,.
\end{equation}

For the scalar field, we take
\begin{equation}
\Phi(r) = \operatorname{arcsinh}\left[\gamma(r)\sqrt{2 + \gamma(r)^2} \right], \quad \text{with} \quad \gamma(r) = \gamma_0 \left( \frac{r_+}{r} \right)^n\,.
\end{equation}
In what follows, we restrict to $n \geq 2$, as this range corresponds to standard boundary conditions  for $\Phi$.

With this choice of scalar profile, we can compute the  action analytically as a function of $\beta$, $y_+$, $\gamma_0$, and $n$. One finds
\begin{multline}
\frac{S_E}{L^2}  = \frac{1}{4} n^2 y_+ \gamma_0^2 \Bigg[ \frac{y_+^2 \tilde{\beta}}{2n - 3} \, {}_2F_1\left(1, 1 - \frac{3}{2n}; 2 - \frac{3}{2n}; -\frac{\gamma_0^2}{2} \right) \\
+ \frac{\tilde{\beta}}{2n - 1} \, {}_2F_1\left(1, 1 - \frac{1}{2n}; 2 - \frac{1}{2n}; -\frac{\gamma_0^2}{2} \right) \\
+ \frac{\tilde{\beta} + 3 y_+^2 \tilde{\beta} - 4\pi y_+}{2n + 1} \, {}_2F_1\left(1, 1 + \frac{1}{2n}; 2 + \frac{1}{2n}; -\frac{\gamma_0^2}{2} \right) \Bigg] \\
+ \frac{y_+}{2} \left( \tilde{\beta} + y_+^2 \tilde{\beta} - 2\pi y_+ + \frac{y_+^2 \tilde{\beta} \gamma_0^2}{3 - 2n} \right) - \frac{1}{2} n y_+ \left( \tilde{\beta} + 2 y_+^2 \tilde{\beta} - 2\pi y_+ \right) \log\left(1 + \frac{\gamma_0^2}{2} \right),
\end{multline}
where ${}_2F_1(a, b; c; z)$ is the Gaussian hypergeometric function, and we defined $y_+ \equiv r_+/L$ and $\tilde{\beta} \equiv \beta / L$.

For fixed $y_+$ and $\tilde{\beta}$, and at large $\gamma_0$, one finds
\begin{equation}
\frac{S_E}{L^2}  \underset{\gamma_0 \to +\infty}{\approx} -\frac{y_+^3 \tilde{\beta} \gamma_0^2}{2 (2n - 3)} + \mathcal{O}(\gamma_0^{3/n})\,,
\label{eq:approx}
\end{equation}
showing that the Euclidean action can become arbitrarily negative as $\gamma_0$ becomes large, even for configurations with constant Ricci scalar. Note that since $n \geq 2$, the term we are neglecting is always subdominant compared to the leading term.

\begin{figure}[tb]
    \centering
    \includegraphics[width=0.8\textwidth]{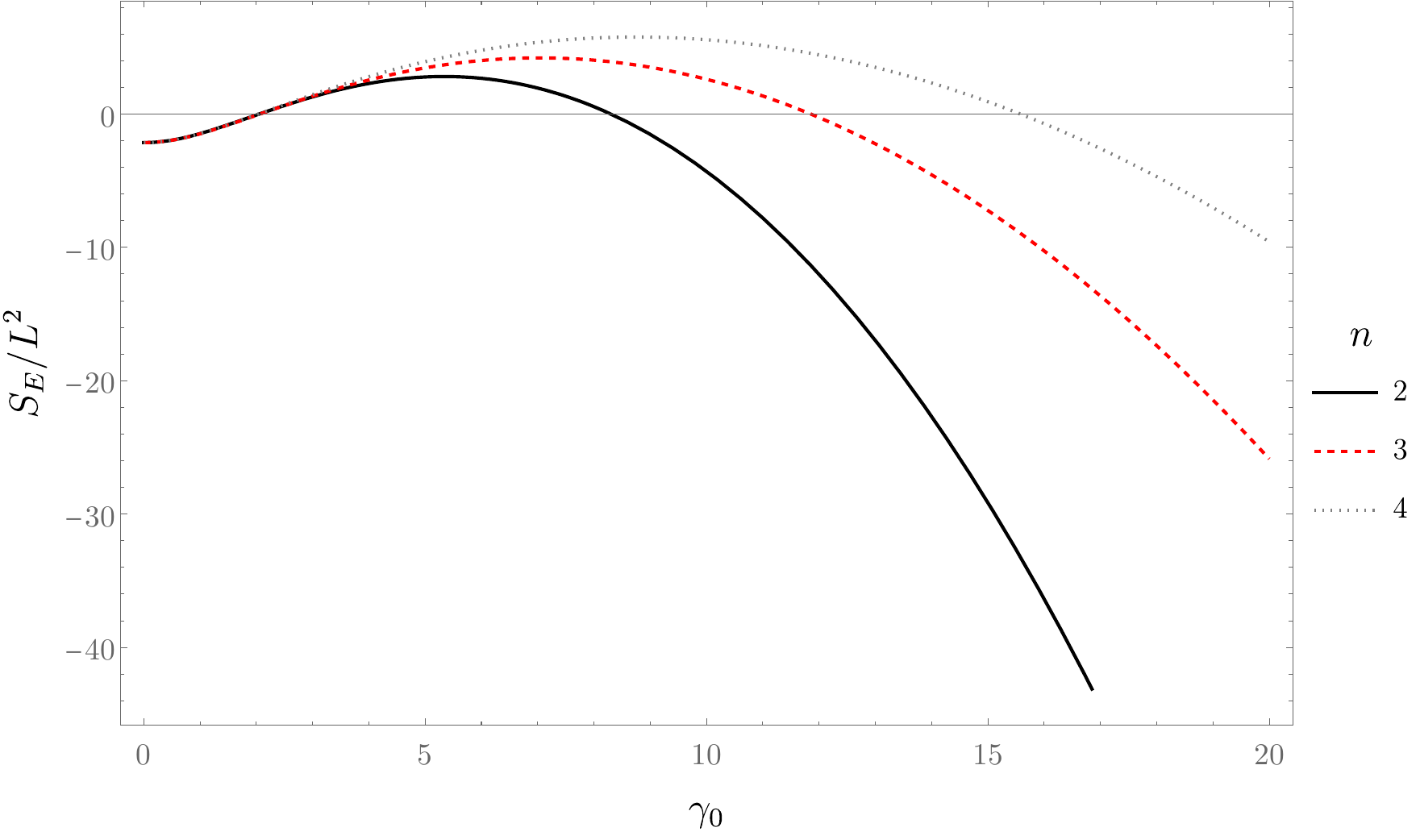}
    \caption{The rescaled Euclidean action $ S_E / L^2$ as a function of $\gamma_0$, for fixed $y_+ = 1$ and $\tilde{\beta} = 1$. As $\gamma_0$ increases, the action becomes increasingly negative, consistent with the analytic approximation in Eq.~(\ref{eq:approx}). Each curve corresponds to a different value of $n$, as indicated in the legend.}
    \label{fig:euclideanconstant}
\end{figure}

In Fig.~\ref{fig:euclideanconstant}, we plot $S_E / L^2$ as a function of $\gamma_0$ for fixed $y_+ = 1$ and $\tilde{\beta} = 1$, for various values of $n$. We clearly see that as $\gamma_0$ becomes large, the Euclidean action becomes arbitrarily negative. The action starts out negative and initially increases with $\gamma_0$, becoming positive over a finite range. However, it eventually turns negative again at larger values of $\gamma_0$. The behavior at large $\gamma_0$ is accurately captured by Eq.~(\ref{eq:approx}). The range over which the action is positive grows with increasing $n$, but the action ultimately becomes negative for all values of $n$.

\section{Attempt at a vacuum example with constant scalar curvature}
\label{sec:complex}

In this section we remove the matter field and try to find a metric with constant scalar curvature such that the vacuum gravitational action (with $\Lambda < 0$) is arbitrarily negative. Once again we work in five dimensions, and adopt the same form of the metric as in Sec. \ref{sec:AdS} with $\Psi(\tau,r)=0$ and $H(r)=1$:
\begin{equation}
{\rm d}s^2=f(r){\rm d}\tau^2+\frac{{\rm d}r^2}{f(r)}
+\frac{r^2}{4}\Big[\sigma_1^2+\sigma_2^2+\left(\sigma_3+2 \Omega(r){\rm d}\tau\right)^2\Big]\,,
\end{equation}
Furthermore, we are interested in geometries with a Euclidean black hole, so that we demand
\begin{equation}
f(r_+)=0\quad\text{and}\quad \frac{f^\prime(r_+)}{4\pi}=\frac{1}{\beta}
\label{eq:BCs}
\end{equation}
for some black hole radius $r_+$. Let us take
\begin{equation}
\Omega(r)=\Omega_0\left(\frac{r_+}{r}\right)^n
\end{equation}
with $n\geq4$. The requirement $n\ge 4$ comes from demanding the usual asymptotics at large $r$, though as an aside we note that the counterterms nevertheless render the action finite for any $n\geq0$. Demanding that we have a constant curvature, i.e.
\begin{equation}
R=-\frac{20}{L^2}
\label{eq:constant}
\end{equation}
gives a second order differential equation that we can readily solve for $f(r)$ with our given boundary conditions:
\begin{multline}
f(r)=\frac{r^2}{L^2}+1+\frac{r_+^2}{r^3} \left[2 r_+-3 r+\frac{r_+^2 \left(4 r_+-5 r\right)}{L^2}+\frac{4 \pi  \left(r-r_+\right) r_+}{\beta }\right]
\\
-\frac{n^2}{4 (n-2) (2 n-5) r^3}\left[(2 n-5) r r_+^4-2 (n-2) r_+^5+r^5
   \left(\frac{r_+}{r}\right){}^{2 n}\right]\Omega _0^2
\end{multline}

We can now go through the holographic renormalisation procedure, using the fact that our geometry has constant curvature, and we find that the Euclidean action is given by
\begin{equation}
S_E=\frac{\pi  \beta r_+^4}{32  L^2} \left(12+\frac{L^2 n^2 \Omega_0^2}{n-2}\right) -\frac{\pi ^2 r_+^3}{2 }+\frac{3 \pi  \beta  r_+^2}{8 }+\frac{3 L^2 \pi  \beta }{32 }
\label{eq:action}
\end{equation}
There are two contributions to the above expression: one coming from the asymptotic behaviour near the boundary, and another term from the horizon. The last term is independent of $r_+$ and corresponds to the Euclidean action of AdS. Now, the reader might wonder why we do not obtain the standard Schwarzschild result when we assume vanishing $\Omega_0$ and the usual dependence of $\beta$ on $r_+$. The answer is that that requiring the scalar curvature to be constant does not require the geometry to agree with standard Schwarzschild, but instaed allows a one parameter generalisation.  This (off-shell) family of geometries allows us to keep the boundary thermal length $\beta$ independent of $r_+$. To recover the usual result we would have to take $\Omega_0=0$ and fix
\begin{equation}
\beta =\frac{2 L^2 \pi  r_+}{L^2+2 r_+^2}\,.
\end{equation}
Indeed, for the choice above and for $\Omega_0=0$ we find
\begin{equation}
S_E-\frac{3 L^2 \pi  \beta }{32 }=\frac{\pi ^2 r_+^3 \left(L^2-r_+^2\right)}{4  \left(L^2+2 r_+^2\right)}
\end{equation}
which shows the usual Hawking-Page transition for $r_+>L$. We note, however, that in this case we are changing the boundary metric (since $\beta$ is dependent on $r_+$). This is not what we want to do.

Returning to Eq.~(\ref{eq:action}), with $\Omega_0\neq0$ and $\beta$ arbitrary, we note that we can choose $\Omega_0$ so that the first term in $S_E$ vanishes. This happens when
\begin{equation}
\Omega_0=\frac{2 \sqrt{3} \sqrt{2-n}}{L n}
\end{equation}
which is purely imaginary for $n>2$ (and in particular for the case $n\ge 4$ required to satisfy the usual asymptotically AdS boundary conditions). With this choice, by taking $r_+$ sufficiently large, we can make the action arbitrarily negative while keeping the boundary length $\beta$ fixed.

Since $\Omega_0$ is imaginary, this metric is complex, so it is not obviously a problem for the Euclidean path integral. However metrics like this often arise  by analytically continuing real Lorentzian rotating solutions, and have played a role in black hole thermodynamics. In particular, it can be shown that our complex spacetime satisfies the KSW condition \cite{Witten:2021nzp,Kontsevich:2021dmb}.  It may thus still be of concern for a proposal to rescale the metric to one of fixed $R$, though we leave investigation of the details for the future.  

\section{Discussion}
\label{sec:disc}

We have shown that imposing the Euclidean constraints on a family of surfaces that foliate the spacetime does not make the Euclidean gravitational action bounded from below.  As noted in the introduction, this also raises questions about the validity of a common method of evaluating the ``stability'' (i.e., the relevance) of Euclidean saddles by studying the action to quadratic order following \cite{Garriga:1997wz,Gratton:1999ya,Kol:2006ga}.

{The particular analyses used in section \ref{sec:compact} and \ref{sec:AdS} were specific to $D=4$ and $D=5$ dimensions respectively.  However, these arguments can be generalized to arbitrary $D\ge 3$ by adding a complex scalar field in analogy with the treatment of section \ref{sec:asymflat}.  We thus expect similar results to also hold in the vacuum case for all $D > 3$, even if they require more work to establish. }

Rescaling to constant scalar curvature is also not satisfactory. It is natural to ask if there are other prescriptions to bound the action. For asymptotically AdS boundary conditions, it is tempting to use gravitational holography, since the dual CFT has a positive definite action. However, holography only equates physical quantities in the two dual theories. It does not map arbitrary off-shell configurations in one theory to off-shell configurations in the other. So one cannot derive a positive gravitational action from the field theory.

When $\Lambda < 0$, our example of an asymptotically AdS space with constant $R < 0$ and arbitrarily negative action included a scalar field with a negative potential coming from supergravity. Since we have not presented an example without the scalar field (and with a real Euclidean metric), one might wonder if there is a theorem ensuring the action is bounded from below in this case. There is indeed such a theorem in $D=2$ and $D=3$ (for $R^3$ topology) 
 \cite{DKM}, but apparently little is known in higher dimensions.  In the presence of matter one could also explore the suggestion of footnote \ref{foot:V} to choose a Weyl rescaling that determines the Ricci scalar in terms of the matter fields.

{The case $D=2$ requires special treatment. Since the Einstein-Hilbert action is a topological invariant, we should consider instead theories of dilaton gravity.  For definiteness, let us consider JT gravity (with arbitrary cosmological constant and, say, with the restriction that the dilaton be non-negative).  If we add a minimally-coupled massless scalar field then, while we have not worked out the details, we expect there to be analogues of all of the constraint-satisfying constructions we presented in sections \ref{sec:compact}-\ref{sec:AdS}.   However, in this case it is known that an alternate prescription due to Saad, Shenker, and Stanford leads to a well-defined ``Euclidean'' path integral in which one integrates the dilaton over the imaginary axis\footnote{There remains a problematic UV divergence in the presence of matter fields like our complex scalar but, as emphasized in section \ref{sec:compact}, such issues are fundamentally different from those considered here.}. This prescription is somewhat analogous to the idea of Wick-rotating the conformal factor in higher dimensions, so its success can be taken as motivation to continue to explore new such scenarios that might be designed to avoid the particular issues raised in section \ref{sec:constR}.}

However, in  the absense of a well defined Euclidean path integral, the best option seems to be to use the Lorentzian path integral as the fundamental starting point as advocated in e.g. \cite{Hartle:2020glw,Schleich:1987fm,Mazur:1989by,Giddings:1989ny,Giddings:1990yj,Marolf:1996gb,Dasgupta:2001ue,Ambjorn:2002gr,Feldbrugge:2017kzv,Feldbrugge:2017fcc,Feldbrugge:2017mbc,Marolf:2020rpm,Colin-Ellerin:2020mva,Colin-Ellerin:2021jev,Marolf:2022ybi}. Euclidean solutions then contribute only if they can be obtained from appropriate deformations of the contour.

\section*{Acknowledgements}

DM thanks Batoul Banihashemi, Jesse Held, Ted Jacobson, Molly Kaplan, and Zhencheng Wang for conversations that led to the start of this project. The work of GH and DM was supported by NSF grant PHY-2408110 and by funds from the University of California. The work of JES was partially supported by STFC consolidated grant ST/X000664/1 and by Hughes Hall College.

\appendix

\section{A positivity result for linearized Euclidean Einstein-Hilbert gravity on the constraint surface}

\label{app:poslin}

This appendix generalizes the argument \cite{Hartle:2020glw,Schleich:1987fm} by Hartle and Schleich for positivity of the linearized Einstein-Hilbert action about flat Euclidean space when the constraints are imposed in the flat slicing.  We also take the opportunity to rewrite the argument in a form that is more transparent for our current purposes.  In particular, 
we show that,  with either asymptotically flat or asymptotically (locally) AdS boundary conditions,  the gravitational action is positive to quadratic order about any real static solution of vacuum Euclidean Einstein-Hilbert gravity (with cosmological constant) when one enforces the constraints defined by the static foliation, and when 
the induced geometry on each slice $\Sigma_\lambda$ is that of an Einstein space of non-positive curvature; i.e., the induced metric on $\Sigma_\lambda$ must satisfy ${\cal R}_{ij} = - \kappa^2 h_{ij}$ for some $\kappa \in {\mathbb R}$.

To begin, we write the Euclidean action for real Euclidean geometries in the canonical first-order form 
\begin{equation}
S_E = \frac{1}{16\pi} \int_{(\tau, x)\in {\mathbb R}\times \Sigma}\left(\tilde \pi^{ij} \dot{h}_{ij} + N\tilde{\cal H} +N^i\tilde {\cal H}_i  \right) {\rm d}\tau\,{\rm d}^d x  + \int_{(\tau, y)\in {\mathbb R}\times \partial \Sigma} {\rm d}\tau\,{\rm d}^{d-1} y\,E_{ADM},
\end{equation}
where $d$ is the dimension of the slice $\Sigma$, $\tilde \pi^{ij}$ is the rank-2 tensor density canonically conjugate to $h_{ij}$ and $E_{ADM}$ is the appropriate ADM boundary term and $N, N^i$ are the Euclidean lapse and shift and $\tilde {\cal H}, \tilde {\cal H}_i$ are the constraints.  The tildes indicate that the constraints have been written as densities on each slice $\Sigma$.

If we expand about a saddle, the linear term in the action will vanish.  Furthermore, the boundary term $E_{ADM}$ for either asymptotically flat or asymptotically AdS boundary conditions is famously precisely linear in the deviation from either empty Euclidean space or empty AdS.  Thus it cannot contribute to the quadratic term in the action.  Noting that $\tilde \pi^{ij}$, $\dot{q}_{ij}$, and the shift $N^i$ all vanish in static backgrounds, and recalling that the constraints vanish in any on-shell background then shows that the second order term will take the form

\begin{eqnarray}
\label{eq:squadcan}
S_E^{(2)} &=& \frac{1}{16\pi} \int_{(\tau, x)\in {\mathbb R}\times \Sigma}\left(\delta \tilde \pi^{ij} \delta \dot{h}_{ij} + \delta N\delta\tilde {\cal H} + N \delta^{(2)}\tilde {\cal H}  +\delta N^i\delta \tilde {\cal H}_i \right)  {\rm d}\tau\,{\rm d}^d x  \, .
\end{eqnarray}

The constraints of the linearized theory are the coefficients  $\delta {\cal H}, \delta{\cal H}_i$ of $\delta N, \delta N_i$.  The linearized momentum constraints $\delta {\cal H}_i=0$ will play no role in our analysis.  Instead, we focus on the linearized Hamiltonian constraint which, since $K_{ij}=0$ in the background, takes the simple form 
\begin{equation}
0 = \delta {\cal H} = \delta\left(-\sqrt{h} {\cal R} + 2 \sqrt{h}\Lambda \right) = \left(\delta \sqrt{h}\right){\RED\cancel{\left(-{\cal R} + 2 \Lambda\right)} } -\sqrt{h} \delta {\cal R},  
\end{equation}
where the factor in red vanishes due to the fact that the background satisfies the constraints.  Our constraint thus reduces to 
\begin{eqnarray}
\label{eq:linconstapp}
0 &=& \delta {\cal R} = {\cal R}_{ij} \delta h^{ij} + h^{ij} \delta {\cal R}_{ij} \cr
&=& -{\cal R}^{ij} \delta h_{ij} + h^{ij}h^{kl} D_kD_{(j} \delta h_{i)l} - h^{ij}h^{kl} D_iD_{j} \delta h_{kl}, 
\end{eqnarray}
where $D_i$ is the covariant derivative on each slice $\Sigma$ induced by the background metric $h_{ij}$ and we have used both $\delta h^{ij} = h^{ik} \delta h_{kl} h^{lj}$ and a standard expression for $\delta {\cal}R_{ij}$ (see e.g. chapter 7 of \cite{Wald:1984rg}).

Now, on each slice we are free to act with a diffeomorphism such that $\delta h_{ij} \rightarrow \delta h_{ij} + D_i \xi_j + D_j \xi_i$ for any vector field $\xi^i$.  We may use this freedom to impose any gauge condition of the form
\begin{equation}
0 =h^{ik} D_k \delta h_{ij} + h^{ik} \alpha D_j \delta h_{ik}  := (L\delta h)_{j}
\end{equation}
with $\alpha > -1/2$ and where the above defines the linear operator $L$ acting on metric perturbations $\delta h_{ij}$.  For such choices of $\alpha$, and for Einstein metrics $h_{ij}$ on $\Sigma$ satisfying our negative curvature condition, the linear operator $\tilde L$ that maps vector fields to vector fields and is defined by $(\tilde L\xi)^i = h^{ij} (L\delta h)_{j}$ for $\delta h_{ij} = D_i \xi_j + D_j \xi_i$ is negative definite (and thus invertible) on the space of square integrable vector fields.   We can therefore use its inverse to impose the above gauge condition on the space of perturbations satisfying asymptotically flat or asymptotically AdS boundary conditions.

Choosing such a gauge makes it straightforward to solve the constraint \eqref{eq:linconstapp}, which (using our condition that ${\cal R}^{ij} = - \kappa^2 h^{ij}$) becomes simply
\begin{equation}
\label{eq:LH2}
0 = \delta {\cal R} = \kappa^2 \delta h  - (\alpha+1) h^{ij} D_i D_j \delta h, 
\end{equation}
with $\delta h := h^{ij} \delta h_{ij}$.   Since $\alpha >-1/2$, the operator $\kappa^2 -(1+\alpha) D^2$ is positive definite and thus invertible.  The constraint thus requires $\delta h=0$ on each slice.  Taking a time derivative of this result then yields $\delta K:= \delta (h^{ij} K_{ij}) =0.$

Recall now that the object that we actually wish to study is the {\it covariant} action defined by integrating out the momenta (i.e., by imposing the equations of motion $0 = \frac{\delta S_E}{\delta \pi^{ij}}$ or, at the linearized level, 
$0 = \frac{\delta S^{(2)}_E}{\delta [\delta \pi^{ij}]}$). Performing this procedure using \eqref{eq:squadcan} and imposing the linearized constraints thus yields precisely
\begin{equation}
\label{eq:LS2final}
S_E^{(2)} = \frac{1}{8\pi}\int_{\Sigma \times {\mathbb R}} \left(K_{ij}K^{ij} - {\RED \cancel{K^2}}\right)\sqrt{h}\,  {\rm d}\tau \,{\rm d}^d x.
\end{equation}
Here the term in red vanishes due to the linearized Hamiltonian constraint \eqref{eq:LH2} and, as noted above, the boundary term $E_{ADM}$ in $S_E$ does not contribute to the quadratic term $S_E^{(2)}.$  What remains in  \eqref{eq:LS2final} is then just the positive-definite kinetic term $K_{ij} K^{ij}$, so 
$S_E^{(2)}$ is non-negative as desired.

\bibliography{action}{}

\providecommand{\href}[2]{#2}\begingroup\raggedright\begin{thebibliography}{10}

\bibitem{Gibbons:1976ue}
G.~W. Gibbons and S.~W. Hawking, ``{Action Integrals and Partition Functions in
  Quantum Gravity},'' \href{http://dx.doi.org/10.1103/PhysRevD.15.2752}{{\em
  Phys. Rev. D} {\bfseries 15} (1977) 2752--2756}.

\bibitem{Coleman:1980aw}
S.~R. Coleman and F.~De~Luccia, ``{Gravitational Effects on and of Vacuum
  Decay},'' \href{http://dx.doi.org/10.1103/PhysRevD.21.3305}{{\em Phys. Rev.
  D} {\bfseries 21} (1980) 3305}.

\bibitem{Witten:1981gj}
E.~Witten, ``{Instability of the Kaluza-Klein Vacuum},''
  \href{http://dx.doi.org/10.1016/0550-3213(82)90007-4}{{\em Nucl. Phys. B}
  {\bfseries 195} (1982) 481--492}.

\bibitem{Penington:2019kki}
G.~Penington, S.~H. Shenker, D.~Stanford, and Z.~Yang, ``{Replica wormholes and
  the black hole interior},''
  \href{http://dx.doi.org/10.1007/JHEP03(2022)205}{{\em JHEP} {\bfseries 03}
  (2022) 205}, \href{http://arxiv.org/abs/1911.11977}{{\ttfamily
  arXiv:1911.11977 [hep-th]}}.

\bibitem{Almheiri:2019qdq}
A.~Almheiri, T.~Hartman, J.~Maldacena, E.~Shaghoulian, and A.~Tajdini,
  ``{Replica Wormholes and the Entropy of Hawking Radiation},''
  \href{http://dx.doi.org/10.1007/JHEP05(2020)013}{{\em JHEP} {\bfseries 05}
  (2020) 013}, \href{http://arxiv.org/abs/1911.12333}{{\ttfamily
  arXiv:1911.12333 [hep-th]}}.

\bibitem{Iliesiu:2020qvm}
L.~V. Iliesiu and G.~J. Turiaci, ``{The statistical mechanics of near-extremal
  black holes},'' \href{http://dx.doi.org/10.1007/JHEP05(2021)145}{{\em JHEP}
  {\bfseries 05} (2021) 145}, \href{http://arxiv.org/abs/2003.02860}{{\ttfamily
  arXiv:2003.02860 [hep-th]}}.

\bibitem{Heydeman:2020hhw}
M.~Heydeman, L.~V. Iliesiu, G.~J. Turiaci, and W.~Zhao, ``{The statistical
  mechanics of near-BPS black holes},''
  \href{http://dx.doi.org/10.1088/1751-8121/ac3be9}{{\em J. Phys. A} {\bfseries
  55} no.~1, (2022) 014004}, \href{http://arxiv.org/abs/2011.01953}{{\ttfamily
  arXiv:2011.01953 [hep-th]}}.

\bibitem{Maldacena:2024spf}
J.~Maldacena, ``{Real observers solving imaginary problems},''
  \href{http://arxiv.org/abs/2412.14014}{{\ttfamily arXiv:2412.14014
  [hep-th]}}.

\bibitem{Ivo:2025yek}
V.~Ivo, J.~Maldacena, and Z.~Sun, ``{Physical instabilities and the phase of
  the Euclidean path integral},''
  \href{http://arxiv.org/abs/2504.00920}{{\ttfamily arXiv:2504.00920
  [hep-th]}}.

\bibitem{Shi:2025amq}
X.~Shi and G.~J. Turiaci, ``{The phase of the gravitational path integral},''
  \href{http://arxiv.org/abs/2504.00900}{{\ttfamily arXiv:2504.00900
  [hep-th]}}.

\bibitem{Gibbons:1978ac}
G.~W. Gibbons, S.~W. Hawking, and M.~J. Perry, ``{Path Integrals and the
  Indefiniteness of the Gravitational Action},''
  \href{http://dx.doi.org/10.1016/0550-3213(78)90161-X}{{\em Nucl. Phys. B}
  {\bfseries 138} (1978) 141--150}.

\bibitem{Hartle:2020glw}
J.~B. Hartle and K.~Schleich, ``{The Conformal Rotation in Linearised
  Gravity},'' in {\em Quantum Field Theory and Quantum Statistics}, C.~J.~I.
  I.~A.~Batalin and G.~A. Vilkovisky, eds., pp.~{67--87}.
\newblock 4, 1987.
\newblock \href{http://arxiv.org/abs/2004.06635}{{\ttfamily arXiv:2004.06635
  [gr-qc]}}.

\bibitem{Schleich:1987fm}
K.~Schleich, ``{Conformal Rotation in Perturbative Gravity},''
  \href{http://dx.doi.org/10.1103/PhysRevD.36.2342}{{\em Phys. Rev. D}
  {\bfseries 36} (1987) 2342--2363}.

\bibitem{Hajicek:1984nb}
P.~Hajicek, ``{Spherically Symmetric Systems of Fields and Black Holes. 3.
  Positivity of Energy and of a New Type Euclidean Action},''
  \href{http://dx.doi.org/10.1103/PhysRevD.30.1185}{{\em Phys. Rev. D}
  {\bfseries 30} (1984) 1185}.

\bibitem{Kuchar:1970mu}
K.~Kuchar, ``{Ground state functional of the linearized gravitational field},''
  \href{http://dx.doi.org/10.1063/1.1665133}{{\em J. Math. Phys.} {\bfseries
  11} (1970) 3322--3334}.

\bibitem{Dasgupta:2001ue}
A.~Dasgupta and R.~Loll, ``{A Proper time cure for the conformal sickness in
  quantum gravity},''
  \href{http://dx.doi.org/10.1016/S0550-3213(01)00227-9}{{\em Nucl. Phys. B}
  {\bfseries 606} (2001) 357--379},
  \href{http://arxiv.org/abs/hep-th/0103186}{{\ttfamily arXiv:hep-th/0103186}}.

\bibitem{Mazur:1989by}
P.~O. Mazur and E.~Mottola, ``{The Gravitational Measure, Solution of the
  Conformal Factor Problem and Stability of the Ground State of Quantum
  Gravity},'' \href{http://dx.doi.org/10.1016/0550-3213(90)90268-I}{{\em Nucl.
  Phys. B} {\bfseries 341} (1990) 187--212}.

\bibitem{Banihashemi:2022htw}
B.~Banihashemi, T.~Jacobson, A.~Svesko, and M.~Visser, ``{The minus sign in the
  first law of de Sitter horizons},''
  \href{http://dx.doi.org/10.1007/JHEP01(2023)054}{{\em JHEP} {\bfseries 01}
  (2023) 054}, \href{http://arxiv.org/abs/2208.11706}{{\ttfamily
  arXiv:2208.11706 [hep-th]}}.

\bibitem{Banihashemi:2024weu}
B.~Banihashemi and T.~Jacobson, ``{The enigmatic gravitational partition
  function},'' \href{http://dx.doi.org/10.1007/s10714-024-03347-0}{{\em Gen.
  Rel. Grav.} {\bfseries 57} no.~2, (2025) 43},
  \href{http://arxiv.org/abs/2411.00267}{{\ttfamily arXiv:2411.00267
  [hep-th]}}.

\bibitem{toappear}
J.~Held, M.~Kaplan, D.~Marolf, and Z.~Wang, ``{title TBD},''. to appear.

\bibitem{Garriga:1997wz}
J.~Garriga, X.~Montes, M.~Sasaki, and T.~Tanaka, ``{Canonical quantization of
  cosmological perturbations in the one-bubble open universe},''
  \href{http://dx.doi.org/10.1016/S0550-3213(97)00780-3}{{\em Nucl. Phys. B}
  {\bfseries 513} (1998) 343--374},
  \href{http://arxiv.org/abs/astro-ph/9706229}{{\ttfamily
  arXiv:astro-ph/9706229}}. [Erratum: Nucl.Phys.B 551, 511--511 (1999)].

\bibitem{Gratton:1999ya}
S.~Gratton and N.~Turok, ``{Cosmological perturbations from the no boundary
  Euclidean path integral},''
  \href{http://dx.doi.org/10.1103/PhysRevD.60.123507}{{\em Phys. Rev. D}
  {\bfseries 60} (1999) 123507},
  \href{http://arxiv.org/abs/astro-ph/9902265}{{\ttfamily
  arXiv:astro-ph/9902265}}.

\bibitem{Kol:2006ga}
B.~Kol, ``{The Power of Action: The Derivation of the Black Hole Negative
  Mode},'' \href{http://dx.doi.org/10.1103/PhysRevD.77.044039}{{\em Phys. Rev.
  D} {\bfseries 77} (2008) 044039},
  \href{http://arxiv.org/abs/hep-th/0608001}{{\ttfamily arXiv:hep-th/0608001}}.

\bibitem{Monteiro:2008wr}
R.~Monteiro and J.~E. Santos, ``{Negative modes and the thermodynamics of
  Reissner-Nordstrom black holes},''
  \href{http://dx.doi.org/10.1103/PhysRevD.79.064006}{{\em Phys. Rev. D}
  {\bfseries 79} (2009) 064006},
  \href{http://arxiv.org/abs/0812.1767}{{\ttfamily arXiv:0812.1767 [gr-qc]}}.

\bibitem{Hertog:2018kbz}
T.~Hertog, B.~Truijen, and T.~Van~Riet, ``{Euclidean axion wormholes have
  multiple negative modes},''
  \href{http://dx.doi.org/10.1103/PhysRevLett.123.081302}{{\em Phys. Rev.
  Lett.} {\bfseries 123} no.~8, (2019) 081302},
  \href{http://arxiv.org/abs/1811.12690}{{\ttfamily arXiv:1811.12690
  [hep-th]}}.

\bibitem{Marolf:2021kjc}
D.~Marolf and J.~E. Santos, ``{AdS Euclidean wormholes},''
  \href{http://dx.doi.org/10.1088/1361-6382/ac2cb7}{{\em Class. Quant. Grav.}
  {\bfseries 38} no.~22, (2021) 224002},
  \href{http://arxiv.org/abs/2101.08875}{{\ttfamily arXiv:2101.08875
  [hep-th]}}.

\bibitem{Loges:2022nuw}
G.~J. Loges, G.~Shiu, and N.~Sudhir, ``{Complex saddles and Euclidean wormholes
  in the Lorentzian path integral},''
  \href{http://dx.doi.org/10.1007/JHEP08(2022)064}{{\em JHEP} {\bfseries 08}
  (2022) 064}, \href{http://arxiv.org/abs/2203.01956}{{\ttfamily
  arXiv:2203.01956 [hep-th]}}.

\bibitem{Loges:2023ypl}
G.~J. Loges, G.~Shiu, and T.~Van~Riet, ``{A 10d construction of Euclidean axion
  wormholes in flat and AdS space},''
  \href{http://dx.doi.org/10.1007/JHEP06(2023)079}{{\em JHEP} {\bfseries 06}
  (2023) 079}, \href{http://arxiv.org/abs/2302.03688}{{\ttfamily
  arXiv:2302.03688 [hep-th]}}.

\bibitem{Aguilar-Gutierrez:2023ril}
S.~E. Aguilar-Gutierrez, T.~Hertog, R.~Tielemans, J.~P. van~der Schaar, and
  T.~Van~Riet, ``{Axion-de Sitter wormholes},''
  \href{http://dx.doi.org/10.1007/JHEP11(2023)225}{{\em JHEP} {\bfseries 11}
  (2023) 225}, \href{http://arxiv.org/abs/2306.13951}{{\ttfamily
  arXiv:2306.13951 [hep-th]}}.

\bibitem{Hertog:2024nys}
T.~Hertog, S.~Maenaut, B.~Missoni, R.~Tielemans, and T.~Van~Riet, ``{Stability
  of axion-saxion wormholes},''
  \href{http://dx.doi.org/10.1007/JHEP11(2024)151}{{\em JHEP} {\bfseries 11}
  (2024) 151}, \href{http://arxiv.org/abs/2405.02072}{{\ttfamily
  arXiv:2405.02072 [hep-th]}}.

\bibitem{SY}
R.~Schoen and S.-T. Yau, ``{Proof of the Positive-Action Conjecture in Quantum
  Relativity},'' \href{http://dx.doi.org/10.1103/PhysRevLett.42.547}{{\em Phys.
  Rev. Lett.} {\bfseries 42} (1979) 547}.

\bibitem{Andersson:1992yk}
L.~Andersson, P.~Chrusciel, and H.~Friedrich, ``{On the Regularity of solutions
  to the Yamabe equation and the existence of smooth hyperboloidal initial data
  for Einsteins field equations},''
  \href{http://dx.doi.org/10.1007/BF02096944}{{\em Commun. Math. Phys.}
  {\bfseries 149} (1992) 587--612}.

\bibitem{allen2022}
P.~T. Allen, J.~M. Lee, and D.~Maxwell, ``Sobolev-class asymptotically
  hyperbolic manifolds and the yamabe problem,''
  \href{http://arxiv.org/abs/2206.12854}{{\ttfamily arXiv:2206.12854
  [math.DG]}}. \url{https://arxiv.org/abs/2206.12854}.

\bibitem{Misner:1973prb}
C.~W. Misner, K.~S. Thorne, and J.~A. Wheeler, {\em {Gravitation}}.
\newblock W. H. Freeman, San Francisco, 1973.

\bibitem{Bianchi:2001de}
M.~Bianchi, D.~Z. Freedman, and K.~Skenderis, ``{How to go with an RG flow},''
  \href{http://dx.doi.org/10.1088/1126-6708/2001/08/041}{{\em JHEP} {\bfseries
  08} (2001) 041},
\href{http://arxiv.org/abs/hep-th/0105276}{{\ttfamily arXiv:hep-th/0105276
  [hep-th]}}.
%%CITATION = HEP-TH/0105276;%%.

\bibitem{Bianchi:2001kw}
M.~Bianchi, D.~Z. Freedman, and K.~Skenderis, ``{Holographic
  renormalization},''
  \href{http://dx.doi.org/10.1016/S0550-3213(02)00179-7}{{\em Nucl. Phys.}
  {\bfseries B631} (2002) 159--194},
\href{http://arxiv.org/abs/hep-th/0112119}{{\ttfamily arXiv:hep-th/0112119
  [hep-th]}}.
%%CITATION = HEP-TH/0112119;%%.

\bibitem{SCH}
R.~Schoen, ``{Conformal deformation of a Riemannian metric to constant scalar
  curvature},'' \href{http://dx.doi.org/10.4310/jdg/1214439291}{{\em J.
  Differential Geom.} {\bfseries 20} (1984) 479--495}.

\bibitem{Cvetic:1999xp}
M.~Cvetic, M.~J. Duff, P.~Hoxha, J.~T. Liu, H.~Lu, J.~X. Lu,
  R.~Martinez-Acosta, C.~N. Pope, H.~Sati, and T.~A. Tran, ``{Embedding AdS
  black holes in ten-dimensions and eleven-dimensions},''
  \href{http://dx.doi.org/10.1016/S0550-3213(99)00419-8}{{\em Nucl. Phys. B}
  {\bfseries 558} (1999) 96--126},
  \href{http://arxiv.org/abs/hep-th/9903214}{{\ttfamily arXiv:hep-th/9903214}}.

\bibitem{Boucher:1984yx}
W.~Boucher, ``{POSITIVE ENERGY WITHOUT SUPERSYMMETRY},''
  \href{http://dx.doi.org/10.1016/0550-3213(84)90394-8}{{\em Nucl. Phys. B}
  {\bfseries 242} (1984) 282--296}.

\bibitem{Townsend:1984iu}
P.~K. Townsend, ``{Positive Energy and the Scalar Potential in Higher
  Dimensional (Super)gravity Theories},''
  \href{http://dx.doi.org/10.1016/0370-2693(84)91610-1}{{\em Phys. Lett. B}
  {\bfseries 148} (1984) 55--59}.

\bibitem{Amsel:2007im}
A.~J. Amsel, T.~Hertog, S.~Hollands, and D.~Marolf, ``{A Tale of two
  superpotentials: Stability and instability in designer gravity},''
  \href{http://dx.doi.org/10.1103/PhysRevD.77.049903}{{\em Phys. Rev. D}
  {\bfseries 75} (2007) 084008},
  \href{http://arxiv.org/abs/hep-th/0701038}{{\ttfamily arXiv:hep-th/0701038}}.
  [Erratum: Phys.Rev.D 77, 049903 (2008)].

\bibitem{Witten:2021nzp}
E.~Witten, ``{A Note On Complex Spacetime Metrics},''
  \href{http://arxiv.org/abs/2111.06514}{{\ttfamily arXiv:2111.06514
  [hep-th]}}.

\bibitem{Kontsevich:2021dmb}
M.~Kontsevich and G.~Segal, ``{Wick Rotation and the Positivity of Energy in
  Quantum Field Theory},'' \href{http://dx.doi.org/10.1093/qmath/haab027}{{\em
  Quart. J. Math. Oxford Ser.} {\bfseries 72} no.~1-2, (2021) 673--699},
  \href{http://arxiv.org/abs/2105.10161}{{\ttfamily arXiv:2105.10161
  [hep-th]}}.

\bibitem{DKM}
M.~Dahl, K.~Kroencke, and S.~McCormick, ``{A volume-renormalized mass for
  asymptotically hyperbolic manifolds},''
  \href{http://arxiv.org/abs/2307.06196}{{\ttfamily arXiv:2307.06196
  [math.DG]}}.

\bibitem{Giddings:1989ny}
S.~B. Giddings, ``{The Conformal Factor and the Cosmological Constant},''
  \href{http://dx.doi.org/10.1142/S0217751X9000163X}{{\em Int. J. Mod. Phys. A}
  {\bfseries 5} (1990) 3811--3830}.

\bibitem{Giddings:1990yj}
S.~B. Giddings, ``{Wormholes, the conformal factor, and the cosmological
  constant},'' in {\em {International Colloquium on Modern Quantum Field
  Theory}}.
\newblock 5, 1990.

\bibitem{Marolf:1996gb}
D.~Marolf, ``{Path integrals and instantons in quantum gravity: Minisuperspace
  models},'' \href{http://dx.doi.org/10.1103/PhysRevD.53.6979}{{\em Phys. Rev.
  D} {\bfseries 53} (1996) 6979--6990},
  \href{http://arxiv.org/abs/gr-qc/9602019}{{\ttfamily arXiv:gr-qc/9602019}}.

\bibitem{Ambjorn:2002gr}
J.~Ambjorn, A.~Dasgupta, J.~Jurkiewicz, and R.~Loll, ``{A Lorentzian cure for
  Euclidean troubles},''
  \href{http://dx.doi.org/10.1016/S0920-5632(01)01903-X}{{\em Nucl. Phys. B
  Proc. Suppl.} {\bfseries 106} (2002) 977--979},
  \href{http://arxiv.org/abs/hep-th/0201104}{{\ttfamily arXiv:hep-th/0201104}}.

\bibitem{Feldbrugge:2017kzv}
J.~Feldbrugge, J.-L. Lehners, and N.~Turok, ``{Lorentzian Quantum Cosmology},''
  \href{http://dx.doi.org/10.1103/PhysRevD.95.103508}{{\em Phys. Rev. D}
  {\bfseries 95} no.~10, (2017) 103508},
  \href{http://arxiv.org/abs/1703.02076}{{\ttfamily arXiv:1703.02076
  [hep-th]}}.

\bibitem{Feldbrugge:2017fcc}
J.~Feldbrugge, J.-L. Lehners, and N.~Turok, ``{No smooth beginning for
  spacetime},'' \href{http://dx.doi.org/10.1103/PhysRevLett.119.171301}{{\em
  Phys. Rev. Lett.} {\bfseries 119} no.~17, (2017) 171301},
  \href{http://arxiv.org/abs/1705.00192}{{\ttfamily arXiv:1705.00192
  [hep-th]}}.

\bibitem{Feldbrugge:2017mbc}
J.~Feldbrugge, J.-L. Lehners, and N.~Turok, ``{No rescue for the no boundary
  proposal: Pointers to the future of quantum cosmology},''
  \href{http://dx.doi.org/10.1103/PhysRevD.97.023509}{{\em Phys. Rev. D}
  {\bfseries 97} no.~2, (2018) 023509},
  \href{http://arxiv.org/abs/1708.05104}{{\ttfamily arXiv:1708.05104
  [hep-th]}}.

\bibitem{Marolf:2020rpm}
D.~Marolf and H.~Maxfield, ``{Observations of Hawking radiation: the Page curve
  and baby universes},'' \href{http://dx.doi.org/10.1007/JHEP04(2021)272}{{\em
  JHEP} {\bfseries 04} (2021) 272},
  \href{http://arxiv.org/abs/2010.06602}{{\ttfamily arXiv:2010.06602
  [hep-th]}}.

\bibitem{Colin-Ellerin:2020mva}
S.~Colin-Ellerin, X.~Dong, D.~Marolf, M.~Rangamani, and Z.~Wang, ``{Real-time
  gravitational replicas: Formalism and a variational principle},''
  \href{http://dx.doi.org/10.1007/JHEP05(2021)117}{{\em JHEP} {\bfseries 05}
  (2021) 117}, \href{http://arxiv.org/abs/2012.00828}{{\ttfamily
  arXiv:2012.00828 [hep-th]}}.

\bibitem{Colin-Ellerin:2021jev}
S.~Colin-Ellerin, X.~Dong, D.~Marolf, M.~Rangamani, and Z.~Wang, ``{Real-time
  gravitational replicas: low dimensional examples},''
  \href{http://dx.doi.org/10.1007/JHEP08(2021)171}{{\em JHEP} {\bfseries 08}
  (2021) 171}, \href{http://arxiv.org/abs/2105.07002}{{\ttfamily
  arXiv:2105.07002 [hep-th]}}.

\bibitem{Marolf:2022ybi}
D.~Marolf, ``{Gravitational thermodynamics without the conformal factor
  problem: partition functions and Euclidean saddles from Lorentzian path
  integrals},'' \href{http://dx.doi.org/10.1007/JHEP07(2022)108}{{\em JHEP}
  {\bfseries 07} (2022) 108}, \href{http://arxiv.org/abs/2203.07421}{{\ttfamily
  arXiv:2203.07421 [hep-th]}}.

\bibitem{Wald:1984rg}
R.~M. Wald,
  \href{http://dx.doi.org/10.7208/chicago/9780226870373.001.0001}{{\em {General
  Relativity}}}.
\newblock Chicago Univ. Pr., Chicago, USA, 1984.

\end{thebibliography}\endgroup
\bibliographystyle{utphys-modified}

\end{document}